\documentclass[longauth]{aa}

\bibpunct{(}{)}{;}{a}{}{,}
\usepackage{amsmath}

\usepackage{graphicx}
\usepackage{txfonts}
\usepackage{hyperref}
\usepackage{natbib}
\hypersetup{colorlinks=true,citecolor=blue}
\usepackage{multicol}
\usepackage{multirow}

\usepackage{caption}
\usepackage{subcaption}

\begin{document}

\title{The VIMOS Public Extragalactic Redshift Survey (VIPERS)
\thanks{Based on observations collected at the European Southern 
Observatory, Cerro Paranal, Chile, using the Very Large Telescope
under programs 182.A-0886 and partly 070.A-9007. Also based on
observations obtained with MegaPrime/MegaCam, a joint project of CFHT
and CEA/DAPNIA, at the Canada-France-Hawaii Telescope (CFHT), which is
operated by the National Research Council (NRC) of Canada, the
Institut National des Sciences de l'Univers of the Centre National de
la Recherche Scientifique (CNRS) of France, and the University of
Hawaii. This work is based in part on data products produced at
TERAPIX and the Canadian Astronomy Data Centre as part of the
Canada-France-Hawaii Telescope Legacy Survey, a collaborative project
of NRC and CNRS. The VIPERS web site is {\tt
http://vipers.inaf.it/}.}}

\subtitle{The coevolution of galaxy morphology and colour to $z\sim 1$}

\author{
J.~Krywult\inst{1} 
\and L.~A.~M.~Tasca\inst{2}
\and A.~Pollo\inst{3,4}
\and D.~Vergani\inst{5}
\and M.~Bolzonella\inst{6}
\and I.~Davidzon\inst{2,6}
\and A.~Iovino\inst{7}
\and A.~Gargiulo\inst{8}
\and C.~P.~Haines\inst{7}
\and M.~Scodeggio\inst{8}
\and L.~Guzzo\inst{7,9} 
\and G.~Zamorani\inst{6}
\and B. Garilli \inst{8}
\and B.~R.~Granett\inst{7}  
\and S.~de la Torre\inst{2}
\and U.~Abbas\inst{10}
\and C.~Adami\inst{2}
\and D.~Bottini\inst{8}
\and A.~Cappi\inst{6}
\and O.~Cucciati\inst{6}       
\and P.~Franzetti\inst{8}
\and A.~Fritz\inst{8}
\and V.~Le Brun\inst{2}
\and O.~Le F\`evre\inst{2}
\and D.~Maccagni\inst{8}
\and K.~Ma{\l}ek\inst{17}
\and F.~Marulli\inst{18,19,6}
\and M.~Polletta\inst{8,24}
\and R.~Tojeiro\inst{22}
\and A.~Zanichelli\inst{23}
\and S.~Arnouts\inst{2}
\and J.~Bel\inst{11}
\and E.~Branchini\inst{12,13,14}
\and G.~De Lucia\inst{16}
\and O.~Ilbert\inst{2}
\and H.~J.~McCracken\inst{20}
\and L.~Moscardini\inst{18,19,6}
\and T.~T.~Takeuchi\inst{21}
}   

\institute{
Institute of Physics, Jan Kochanowski University, ul. Swietokrzyska 15, 25-406 Kielce, Poland 
\and Aix Marseille Universit\'e, CNRS, LAM (Laboratoire d'Astrophysique de Marseille) UMR 7326, 13388, Marseille, France  
\and Astronomical Observatory of the Jagiellonian University, Orla 171, 30-001 Cracow, Poland 
\and National Centre for Nuclear Research, ul. Hoza 69, 00-681 Warszawa, Poland 
\and INAF - Istituto di Astrofisica Spaziale e Fisica Cosmica Bologna, via Gobetti 101, I-40129 Bologna, Italy 
\and INAF - Osservatorio Astronomico di Bologna, via Ranzani 1, I-40127, Bologna, Italy 
\and INAF - Osservatorio Astronomico di Brera, Via Brera 28, 20122 Milano, via E. Bianchi 46, 23807 Merate, Italy 
\and INAF - Istituto di Astrofisica Spaziale e Fisica Cosmica Milano, via Bassini 15, 20133 Milano, Italy 
\and Dipartimento di Fisica, Universit\`a di Milano-Bicocca, P.zza della Scienza 3, I-20126 Milano, Italy 
\and INAF - Osservatorio Astrofisico di Torino, 10025 Pino Torinese, Italy 
\and Aix-Marseille Universit\'e, CNRS, CPT (Centre de Physique  Th\'eorique) UMR 7332, F-13288 Marseille, France 
\and Dipartimento di Matematica e Fisica, Universit\`{a} degli Studi Roma Tre, via della Vasca Navale 84, 00146 Roma, Italy 
\and INFN, Sezione di Roma Tre, via della Vasca Navale 84, I-00146 Roma, Italy 
\and INAF - Osservatorio Astronomico di Roma, via Frascati 33, I-00040 Monte Porzio Catone (RM), Italy 
\and Department of Astronomy, University of Geneva, ch. d’Écogia 16, CH-1290 Versoix, Switzerland 
\and INAF - Osservatorio Astronomico di Trieste, via G. B. Tiepolo 11, 34143 Trieste, Italy 
\and National   Centre   for   Nuclear   Research,   ul.   Hoza   69,   00-681, Warszawa, Poland 
\and Dipartimento di Fisica e Astronomia - Universit\`{a} di Bologna, viale Berti Pichat 6/2, I-40127 Bologna, Italy 
\and INFN, Sezione di Bologna, viale Berti Pichat 6/2, I-40127 Bologna, Italy 
\and Institute d'Astrophysique de Paris, UMR7095 CNRS, Universit\'{e} Pierre et Marie Curie, 98 bis Boulevard Arago, 75014 Paris, France 
\and Division of Particle and Astrophysical Science, Nagoya University, Furo-cho, Chikusa-ku, Nagoya 464-8602, Japan 
\and Institute of Cosmology and Gravitation, Dennis Sciama Building, University of Portsmouth, Burnaby Road, Portsmouth, PO1 3FX 
\and INAF - Istituto di Radioastronomia, via Gobetti 101, I-40129, Bologna, Italy 
\and IRAP, 9 av. du colonel Roche, BP 44346, F-31028 Toulouse cedex 4, France  
}

\offprints{J. Krywult, \email{krywult@ujk.edu.pl}}
\date{Received --/Accepted --}

\abstract
{Understanding how galaxy properties evolve is one of the challenges of
extragalactic astrophysics. One possible approach is to statistically
analyse large samples of galaxies: to this aim galaxies must be
separated in different classes sharing the same characteristics.
Therefore, the study of the separation of galaxy types and of the
evolution of the specific parameters used in the classification are
fundamental to understand galaxy evolution.}
{We explore the evolution of the statistical distribution of galaxy
morphological properties and colours over the redshift range $0.5 < z <
1$, combining high-quality imaging data from the CFHT Legacy Survey with
the large number of redshifts and extended photometry from the VIPERS
survey.
}
{Galaxy structural parameters are measured by fitting S\'ersic profiles
to $i$-band images and then combined with absolute magnitudes, colours
and redshifts, to trace the evolution in a multi-parameter space. We
analyse, using a new method, the combination of colours and structural
parameters of early- and late-type galaxies in luminosity--redshift
space.}
{We found that both the rest-frame colour distributions in the (U-B) vs.
(B-V) plane and the S\'ersic index distributions are well fitted by a
sum of two Gaussians, with a remarkable consistency of red-spheroidal
and blue-disky galaxy populations, over the explored redshift
($0.5<z<1$) and luminosity ($-1.5<B-B_*<1.0$) ranges.
The combination of the UBV rest-frame colour and S\'ersic index $n$ as a
function of redshift and luminosity allows us to present the structure of
early- and late-type galaxies and their evolution.
We found that early type galaxies display only a slow change of their
concentrations since $z\sim1$; it is already established by $z\sim1$ and
depends much more strongly on their luminosities. In contrast, late-type
galaxies get clearly more concentrated with cosmic time since $z\sim1$,
with only little evolution in colour, which remains dependent mainly on
their luminosity. This flipped luminosity (mass) and redshift dependence
likely reflects different evolutionary tracks of early- and late-type
galaxies before and after $z\sim1$.
}
{
The combination of rest-frame colours and S\'ersic index $n$ as a
function of redshift and luminosity leads to a precise statistical
description of the structure of galaxies and their evolution.
Additionally, the proposed method provides a robust way to split
galaxies into early and late types.
}


\authorrunning{J. Krywult et al.}
\titlerunning{VIPERS: evolution of shape and colour bimodalities}

\maketitle


\section{Introduction}
\label{sect_intro}

The human eye and brain have evolved to be able to rapidly pick up
underlying similarities and subtle differences amongst a set of
objects (even unconsciously) allowing them to be efficiently and
reliably identified and ordered into categories.  As for galaxy
studies it is common use to divide sources into populations according
to specific galaxy properties.  \citet{Hubble1926} provides the first
statistical classification of extragalactic nebulae based on their
shapes.  Since then the Hubble tuning fork has been used to divide
galaxies into ellipticals, spirals and irregulars, with various
degrees of complexity and details. The original classification scheme
went through various modifications and found its definitive exposition
in \citet{Sandage1961}.  Still, the well-defined galaxy segregation
observed in the local universe starts to lose its discriminatory power
when moving to higher redshifts where galaxies have more irregular and
diverse shapes, and new classification schemes should be introduced
\citep[e.g.][]{vanderWel2007,Kartaltepe2015}.

Due to the impressive amount of photometric data produced by large
galaxy surveys, it is necessary to move from human classifiers to
automatic techniques.  The standard approach is to identify a series
of parameters which correlate with the visual morphology of a galaxy
and to define the parameter--space which best identifies a specific
morphological type \citep[e.g.][]{Abraham1996,Conselice2000,Lotz2008}.
Among the non-parametric diagnostics of galaxy structure the more
traditionally used are galaxy asymmetry, concentration, Gini
coefficient, the 2nd-order moment of the brightest $20\%$ of galaxy
pixels, clumpiness (or smoothness) and ellipticity
\citep{Abraham2003,Lotz2004}.  A widely used parametric description of
the galaxy light profile is based on the exponent of the S\'ersic law
fit to the galaxy surface brightness distribution \citep{Sersic1963}.
The S\'ersic index $n$, that quantifies the concentration of the
light, has been commonly used as a selection criterion to divide early
and late--type galaxies in many investigations (e.g.
\citealt{Driver2006} applied $n=2$ to the galaxies from Millennium
Galaxy Catalogue; \citealt{Cassata2011} used $n=2$ on the high-$z$ HST
galaxies).  \citet{Ravindranath2004} analysed a sample of nearby
galaxies with visual morphologies determined by~\citet{Frei1996}
artificially redshifted to  $z=0.5$ and $1.0$, and found that the
single S\'ersic profile index $n=2$ well separates early- and
late-type galaxies, even in the presence of dust or star forming
regions.
 
Alongside the rather qualitative classification criteria at the basis
of the Hubble-Sandage system, a more quantitative interpretation
related to how physical parameters (e.g. stellar mass, specific
angular momentum, ages, cold gas fraction, etc.) vary along the Hubble
sequence, can be developed \citep[see][for an extensive
review]{Roberts1994}.  Hubble's early-type galaxies (ellipticals and
lenticulars) are usually redder in optical colours, more luminous and
massive, with older stellar populations, and smaller reservoirs of gas
and dust; conversely, late-type galaxies (spirals and irregular
galaxies) are generally less massive, show younger stellar populations
and have bluer colours~\citep[e.g.][]{deVaucouleurs1961,Roberts1994,
Kennicutt1998,Bell2004,Bundy2005, Bundy2006,Haynes1984,Noordermeer2005}.
Many studies suggest that these correlations hold at least up to
$z\sim1$
\citep{Fritz2009a,Fritz2009b,Pozzetti2010,Bolzonella2010,Kovac2010,Tasca2009}.
In particular, the morphology-colour correlation is traced back to at
least up to $z\sim 2$ \citep[e.g.][]{Bassett2013}.

Similarly to what is seen in the distribution of morphological types,
galaxy rest-frame colours tend to segregate into a bimodal
distribution. This is best evidenced by the colour--magnitude (or
colour--stellar mass) diagram, in which two clear loci are
preferentially occupied by the blue and the red populations, known
respectively as the ``blue cloud'' (or sometimes ``blue sequence'')
and the ``red sequence''.  Galaxy colours reflect the ages and star
formation histories of the mean galaxy stellar population. To
understand the origin of the observed colour bimodality would
therefore help to shed light on the main galaxy evolution mechanisms
at play and their relative timescales.  It is now commonly
accepted that the total stellar mass within the blue cloud shows very
little growth since $z\sim1$, while the red sequence has grown by at
least a factor $\sim2$ \citep[e.g.][]{Cimatti2006,Arnouts2007}.  The
most popular scenario invoked to explain the growth of red galaxies is
a migration of a significant fraction of star-forming systems from the
blue cloud to the red sequence, due to different quenching processes.
Observational studies of high-mass (central) galaxies prefer a
self-regulated mass quenching, while quenching in low-mass (satellite)
galaxies has likely been mainly due to environmental and/or merging
influences \citep[e.g.][]{Peng2010,Peng2012,Wetzel2014}.

When a narrow luminosity bin is considered, the resulting distribution
of colours can be described fairly well as a sum of two Gauss
functions, although it has also been shown that an additional,
intermediate population, inhabiting the so-called ``green valley''
between the two main sequences, may also be required
\citep{Wyder2007,Mendez2011,Schawinski2009,Coppa2011,Loh2010,Lackner2012,Brammer2009}.
These objects are commonly thought to represent a transition phase
from the blue cloud to the red sequence, showing the star formation
quenching mechanism at work~\citep{Pozzetti2010}.  \citet{Arnouts2013}
found that actively star--forming and quiescent galaxies segregate
themselves particularly well in the $NUV-r$ versus $r-K$ plane.  More
recently \citet{Moutard2016}, using the multi--wavelength information
collected in the VIPERS region, reported a locus in the $NUVrK$
diagram inhabited by massive galaxies with a variety of morphologies
probably transiting from the star forming to the quiescent
populations.  A similar behavior is observed out to
$z=1.3$~\citep{Coppa2011} and the ``green valley'' population is still
present when using different rest-frame colours, such as
$U-B$~\citep{Nandra2007,Yan2011}, $U-V$
\citep{Brammer2009,Moresco2010} and $NUV-r$
\citep{Wyder2007,Fritz2014}.
  
Understanding the physical processes responsible for the observed
bimodality in morphology and colour and its dependence on the galaxy
environment is a major challenge in the field of galaxy
evolution~\citep[e.g.][]{Tasca2009}.  To shed some light on how the
progenitors of galaxies in the local universe have acquired their
shapes and physical properties, large surveys, as well as the
classification of galaxies at different epochs, are needed.  The VIMOS
Public Extragalactic Survey \citep[VIPERS;][]{Guzzo2013} fulfills
these requirements over the redshift range $0.5<z<1.2$. VIPERS is a
spectroscopic redshift survey which provides on one side a unique
combination of volume and density, and on the other side excellent
5--band photometric coverage with the Canada--France--Hawaii Telescope
Legacy Survey Wide (CFHTLS-Wide), suitable to obtain galaxy
morphologies, colours and rest-frame spectral energy distributions
(SEDs), from which physical properties such as stellar mass can be
derived \citep[e.g.][]{Fritz2014}.

The purpose of this work is to develop a robust method to classify
galaxies from intermediate redshift range in order to analyse their
colour and morphological observational parameters from ground-based
observations.

This paper is organized as follows.
In Sect.~\ref{sect_data} we summarize the data used.
In Sect.~\ref{sect_rest_frame_colour}, we describe the method of
bimodality analysis using galaxy colour and redshift and discuss the
evolutionary trends in colour bimodality.
In Sect.~\ref{sect_sersic_index} we present the methodology of
measurement of S\'ersic parameters of the VIPERS galaxies from the
CFHTLS images, discuss the bimodality of the S\'ersic index distribution
and present evolutionary effects on the S\'ersic index.
In Sect.~\ref{sect_literat} we compare our results with the published
relations involving the measurement of the S\'ersic index.
In Sect.~\ref{sect_n_col} we introduce a new method to classify
galaxies, fully exploiting the 2D distribution in the colour-shape plane
as a function of rest-frame magnitude and redshift.
In Sect.~\ref{sect_n_col_evol} we discuss the implication for the
evolution of early- and late-type galaxies of this new classification
scheme and we summarise our results in Sect.~\ref{sect_concl}.
In Appendix.~\ref{appendix_tests} and~\ref{appendix_psf} we show
the tests of reliability of the S\'ersic function profile fitting
procedure.

For clarity, for the remainder of this article when describing the two
main galaxy populations we will call them red and blue when they have
been selected simply according to their colours, spheroid-like and
disc-like when selected solely based upon their S\'ersic index, and
early-type and late-type when the populations are selected from both
colour and morphology.

In our analysis all magnitudes are given in the AB photometric system.
Throughout the cosmological model with a matter density parameter
$\Omega_m = 0.3$, cosmological constant density parameter
$\Omega_\Lambda = 0.7$ and Hubble constant \mbox{${\rm H}_0 = 70\ {\rm
km\, s}^{-1}\,{\rm Mpc}^{-1}$} is assumed.


\section{Data}
\label{sect_data}

\subsection{The VIPERS project}

The VIMOS Public Extragalactic Redshift Survey (VIPERS) is an
ESO Large Programme aimed at measuring redshifts for
$\sim 10^5$ galaxies at $0.5 < z \lesssim 1.2$, to accurately and
robustly measure clustering, the growth of structure (through
redshift-space distortions), and galaxy properties at an epoch when
the Universe was about half its current age.  Spectroscopic targets
were first selected to a limit of $i<22.5$ in two fields (namely
W1 and W4) of the Canada-France-Hawaii Telescope Legacy Survey Wide
\citep[CFHTLS T0005 release,][]{Mellier2008}, further applying a
simple and robust $gri$ colour pre-selection to effectively remove
galaxies at $z<0.5$.  Spectra have been observed with the VIMOS
multi-object spectrograph~\citep{LeFevre2003} at moderate resolution
($R=210$) using the LR Red grism. This provides a wavelength coverage
of $5500-9500$\,\AA\ and a typical radial velocity error of
$141\,\mathrm{km}\, \mathrm{s}^{-1}$.  Coupled to the
``short-slits'' observing strategy described in \citet{Scodeggio2009},
the colour pre-selection allows us to double the galaxy sampling rate
(which is $\sim 40$\% in the redshift range of interest) with respect
to a pure magnitude-limited sample.

At the same time, the total area (about $24\,\mathrm{deg}^2$) and the
depth of VIPERS result in a large volume, $5\times 10^{7}$\,
h$^{-3}$\, Mpc$^{3}$, analogous to that of the local 2dFGRS.  Such
combination of sampling and depth is unique among current redshift
surveys at~$z>0.5$.  Further details on the design of VIPERS, along
with its data products, can be found in \citet{Guzzo2014}.

In the present paper, we investigate the morphological properties of
galaxies in the VIPERS Public Data Release 1 \citep[PDR-1,
see][]{Garilli2014}, and their interplay with rest-frame colours.
This catalogue\footnote{The PDR-1 catalogue is fully available to the
public through the official website \url{http://vipers.inaf.it}}
includes $55\,358$ galaxies with spectroscopic redshifts
($z_\mathrm{spec}$) over about $10\,\mathrm{deg}^2$.

Besides the spectroscopic redshift, each galaxy in the PDR-1 catalogue
is provided with $u,g,r,i,z$ apparent magnitudes, as estimated by the
Terapix team using \texttt{SExtractor} \citep{Bertin1996}.  These
(MAG\_AUTO) magnitudes are part of the CFHTLS-T0005 data release and
were derived in double image mode in order to match the same aperture
in all bands.

\subsection{Photometric data}

From the PDR-1 catalogue we selected only galaxies with redshifts
measured with the highest reliability, i.e.~with quality flag
$z_\mathrm{flag}=[2,3,4,9]$ according to the classification presented
in~\citet{Guzzo2014} \footnote{The same flag scheme was used in
previous spectroscopic surveys as VVDS \citep{LeFevre2013} and
zCOSMOS \citep{Lilly2007}.}.  Moreover, due to small numbers of high-redshift
galaxies we restrict our analysis to $z_\mathrm{spec}\le 0.95$,
reducing the samples to $20\,208$ and $18\,299$ galaxies in the
W1 and W4 fields respectively.

All spectrophotometric rest-frame properties of the VIPERS galaxies
were derived using the SED fitting program Hyperzmass
\citep{Bolzonella2010}. Absolute magnitudes were derived using the
apparent magnitude that most closely resembles the observed
photometric passband, shifted to the redshift of the galaxy under
consideration, before applying colour and k-corrections derived from
the best-fit SED \citep[see details in][]{Fritz2014}.  To investigate
the dependence of morphology and colour of galaxies on their redshift,
we corrected their absolute magnitudes to account for their intrinsic
evolution, as derived from the characteristic luminosity parameter
($L^{*}$) of the luminosity function (LF) in the \citet{Schechter1976}'s
equation.  For this purpose, we used the global $B$-band LF in the
redshift range from $z=0.5$ to $1.3$ presented in Table~3 of
\citet{Fritz2014}.
These data have been used to compute the linear
approximation of evolution with redshift of the characteristic magnitude
$B_{\rm ev}$ and to define the $\Delta B_{\rm ev}$ luminosity
by the equation:
\begin{equation}
\Delta B_{\rm ev} = M_{B}-B_{\rm ev}(z) = M_{B}+19.90+1.59z \;.
\label{Bstar_eq}
\end{equation}
Considering the evolution of the whole galaxy population,
without division them into the blue and red populations, we found a
slightly steeper $B_{\rm ev}$ evolution than reported in other
studies~\citep[e.g.][]{Faber2007}.
They found that in the redshift range $0<z<1$ the characteristic
magnitude $B_{\rm ev}$ evolves in $z$ with a slope $-1.23 \pm 0.29$,
whereas our study gives $-1.59 \pm 0.20$.
However, the results are consistent within $1\sigma$~uncertainties.

\begin{figure}
\centering
\includegraphics[width=\hsize]{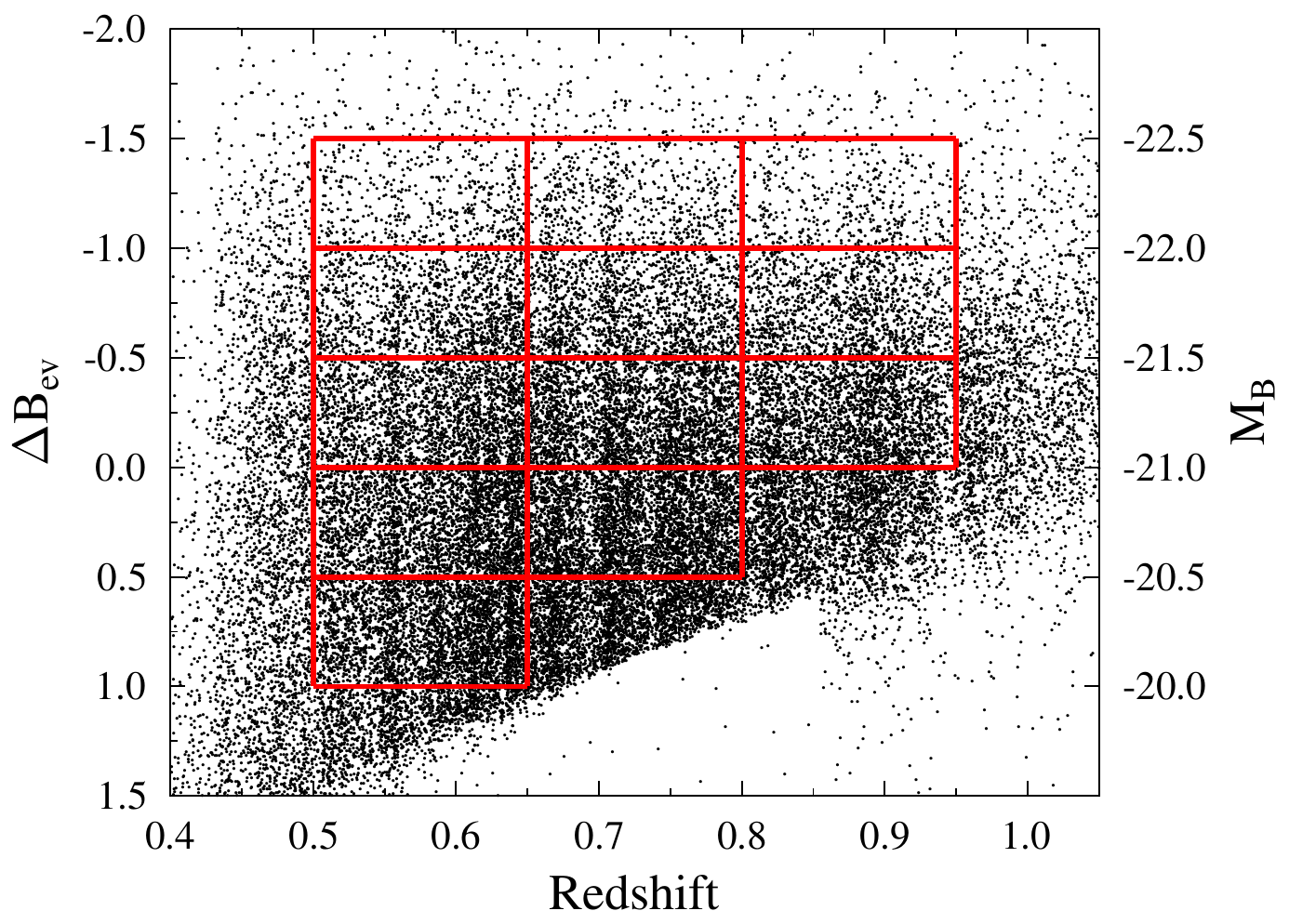}
\caption{
Distribution of $\Delta B_{\rm ev}$ as defined in Eq.~\ref{Bstar_eq} for galaxies
in the VIPERS sample as a function of redshift $z$. The red lines
enclose the selected sub-samples of galaxies.
The right-side vertical axis shows the values of the absolute magnitude
$M_B$ for a fixed redshift $z=0.7$.}
\label{B-Bstar_all}
\end{figure}

In Figure~\ref{B-Bstar_all} we present the distribution of rest-frame
$\Delta B_{\rm ev}$ as a function of redshift for the VIPERS galaxies.
The left-side vertical axis shows the $\Delta B_{\rm ev}$ value, whereas
the right-side one gives the absolute magnitude $M_B$ at the mean VIPERS
redshift $z=0.7$.
As expected, due to selection effects, we progressively lose the faint
population to higher redshifts, leaving only the brighter objects.
In the present study we considered 12 volume limited subsamples
represented by the red boxes in Fig.~\ref{B-Bstar_all}.
Each subsample is statistically complete, spans \mbox{$\Delta B_{\rm ev}=0.5$}
magnitudes and a redshift range $\Delta z=0.15$.

\subsection{CFHTLS imaging}

The morphological analysis has been based on the study of the 2D
surface brightness profile of the VIPERS galaxies.  To model the light
profile of galaxies in the VIPERS PDR-1 we use CCD images in the
$i$-band from eighteen W1 and eleven W4 CFHTLS fields covering
$28\,{\rm deg}^2$ of the VIPERS project.  While the VIPERS PDR-1
catalogue is based on the Terapix T0005 release, for the analysis of
the structural parameters we use a more recent version of the CFHTLS
data \citep[i.e. T0006,][]{Goranova2009}.  A full description of the
CFHTLS data processing including calibration, stacking and mosaicing
is provided in \citet{Mellier2008} and \citet{Goranova2009}.  The
public data from Terapix T0006 are organised in $1\degr\times 1\degr$
fields and have a pixel scale of $0.186\arcsec$.  The mean seeing,
i.e. full width at half maximum (FWHM) of stellar sources, depends on
the filter of the CFHTLS images and is equal to $0.85\arcsec$,
$0.78\arcsec$, $0.72\arcsec$, $0.64\arcsec$, $0.68\arcsec$ in the
$u,g,r,i/y$ (the filter $i$ broke in 2006 and it was replaced by a
similar, but not identical, filter, called $y$) and $z$-bands,
respectively~\citep{Goranova2009}.

To secure the quality of the derived morphological parameters, we used
CCD tiles in the $i$ photometric band where the mean FWHM is smallest.
Objects were extracted by independently running \texttt{SExtractor} on
the CFHTLS tiles in the T0006 release.  This means that the centroid
of photometric sources can be slightly different from the coordinates
of the corresponding VIPERS spectroscopic objects.  We associated
spectroscopic and photometric sources on the basis of their relative
(projected) distance, assuming a maximum matching radius equal to
$1\arcsec$.  For $98.6$\% of the objects the distance between the
VIPERS galaxy (i.e., its position according to T0005) and the one in
the T0006 release is less than $0.3\arcsec$, and for only $0.3$\%
objects it is larger than $0.5\arcsec$.  Objects with distance larger
than~$1\arcsec$ were excluded from the present analysis.


\section{Rest-frame colours}
\label{sect_rest_frame_colour}

\subsection{Galaxy colour -- based classification}
\label{sect_col}

To probe the colour distribution of VIPERS galaxies we use the
rest-frame $(U-B)$ versus $(B-V)$ colour-colour plot, based on the
absolute magnitudes derived in~\cite{Fritz2014}.

\begin{figure}
\centering
\includegraphics[width=\hsize]{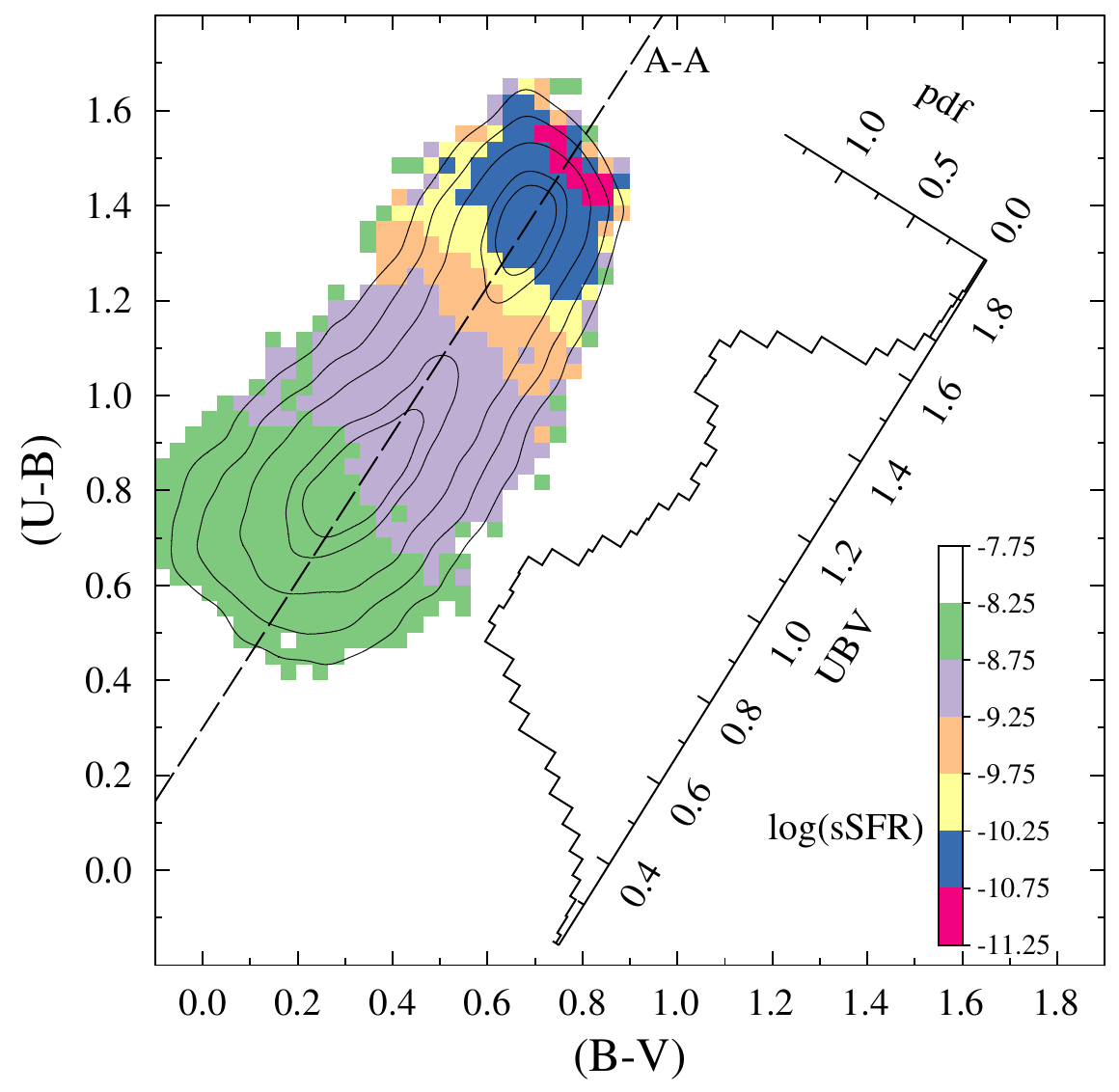}
\caption{Density of the VIPERS galaxies in the rest-frame $(U-B)$
versus $(B-V)$ colour-colour diagram. The contour lines show the
galaxy density distribution in five equally spaced levels from
$10$\% to $99$\% of the maximum value.  The histogram shows the
galaxy number density distribution projected along the line
\mbox{$A-A$} connecting the two maxima of this distribution. The
colours show the median sSFR 1/yr values of galaxies derived from SED
fitting in seven equally spaced logarithmic bins.}
\label{BV_UB}
\end{figure}

The isodensity contour lines presented in Fig.~\ref{BV_UB} show an
evident bimodality in the rest-frame colours, with two well separated
peaks.
We define the combined colour $UBV$ by projecting the galaxy rest-frame
colours along the \mbox{$A-A$} dashed line that connects the two
density peaks of Fig.~\ref{BV_UB}.
In this way the separation of the red and blue populations is even more
prominent than using the one-dimensional analysis, i.e. based only on
$(U-B)$ or $(B-V)$ rest-frame colours.
The dashed line that defines the combined $UBV$ rest-frame colour is
described by the following equation:
\begin{equation}
UBV = (B-V) \times \cos(\theta) - (U-B) \times \sin(\theta),\\
\label{UB_BV_rot_eq}
\end{equation}
where $\theta=58.08\degr$ is the slope of the $A-A$ line.
The $UBV$ rest-frame colour separation of the two peaks along the
the $UBV$ line is equal to 0.71, to be compared with 0.61 and 0.37 when
it is projected on the $(U-B)$ and $(B-V)$ axes, respectively.

Figure~\ref{BV_UB} is colour coded by the median specific Star Formation
Rate (sSFR is defined as the star formation rate per unit stellar mass
of a galaxy) of galaxies inside a given small range of $(U-V)$ and
$(B-V)$ colours.
The sSFRs are derived via SED fitting.
Values of constant sSFR are almost perpendicular to the line connecting
the two colours peaks ($A-A$ line), with values of sSFR steadily
decreasing with $UBV$ rest-frame colour along the line $A-A$.
The correlation between the $UBV$ colours and sSFR is therefore clearly
evident, with blue colours corresponding to higher values of sSFR and
red galaxies being mostly quiescent.
It is also noticeable how this correlation is stronger than the one with
$(U-B)$ or $(B-V)$ colours used independently.
The local minimum of the $UBV$ probability distribution function
corresponds to a value of sSFR~$\approx 10^{-10}\,{\rm yr}^{-1}$, which
is the value adopted by \citet{Davidzon2016} to separate active and
passive galaxies at these redshifts.
Therefore, even if in the following analysis we will use the colour
$UBV$, we should keep in mind that this parameter can be considered a
good proxy of sSFR.


\subsection{Galaxy colour bimodality}
\label{sect_bimod_c}

\begin{figure*}
\centering
\includegraphics[width=\hsize]{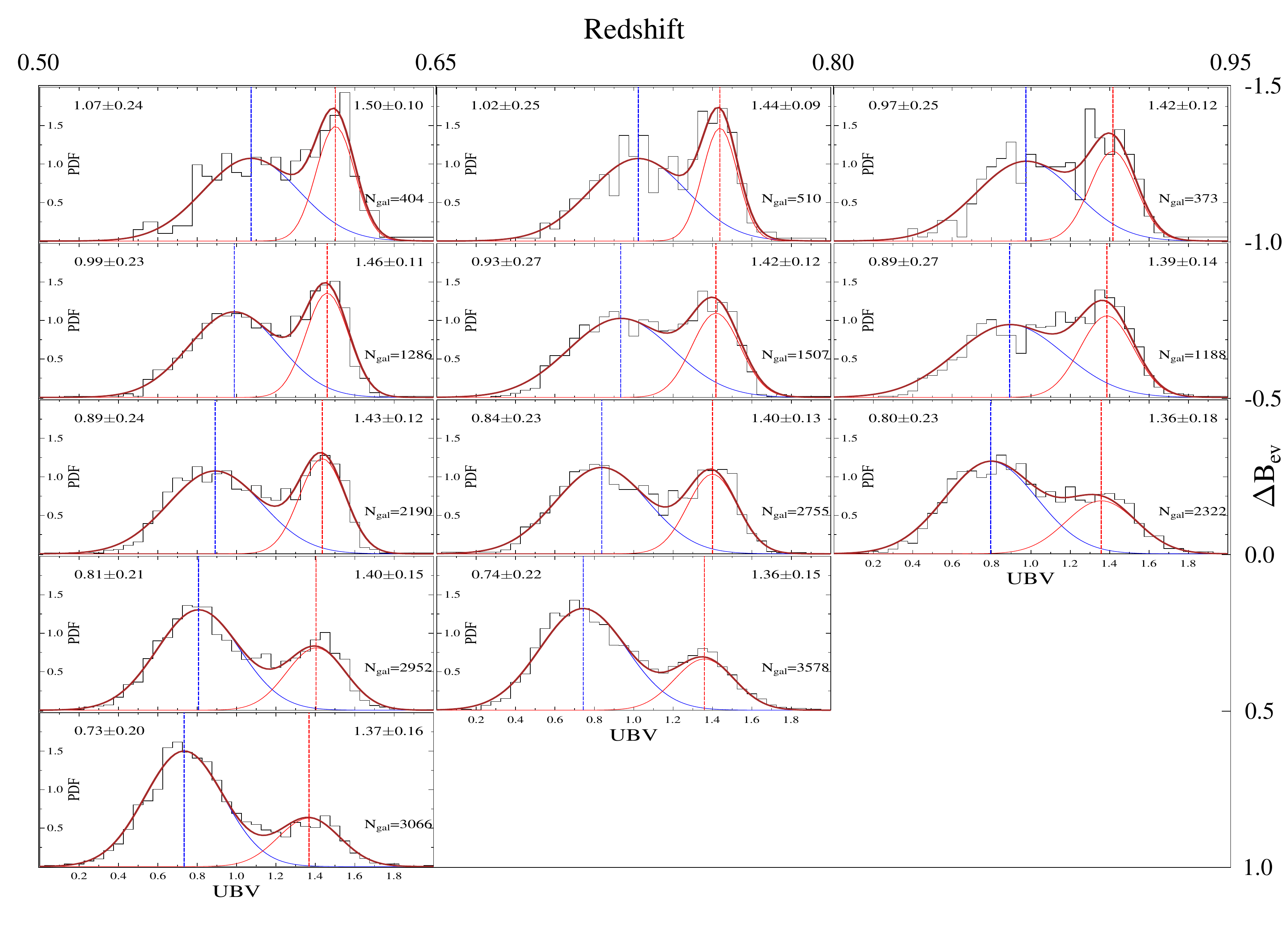}
\caption{$UBV$ rest-frame colour distributions (black histograms) of
VIPERS galaxies in different redshift (increasing from left to right)
and luminosity (from top to bottom) bins.
The blue and red curves represent the Gaussian components fitting the
colour distribution of the two galaxy populations, and the vertical
dashed lines mark the maxima of the Gauss functions.
The solid brown line shows the sum of the two Gaussians.
The central values and $1\sigma$ widths of the Gaussians for the blue
and red galaxy populations are labeled in each panel, in the top left and
right respectively.
The number of galaxies considered in each bin is also shown in the
bottom right of each panel.}
\label{gv_B_grid}
\end{figure*}

To investigate the dependence of the $UBV$ rest-frame bimodality on
galaxy luminosity and redshift we have computed the distribution of the
combined rest-frame colour $UBV$ defined in Eq.~\ref{UB_BV_rot_eq} in each of the
subsamples shown in Fig.~\ref{B-Bstar_all}, i.e. five equally-sized bins
in $\Delta B_{\rm ev}$ of width 0.5\,mag and three bins in redshift each
of width 0.15 in $z$ ($0.50<z\le0.65$, $0.65<z\le0.80$ and
$0.80<z\le0.95$).
The results are presented in Fig.~\ref{gv_B_grid}.
The bimodality is a persistent feature over the whole
luminosity-redshift range explored. The shape of the PDF, however
changes.
The red population (the red line) is dominant at bright luminosities,
whereas the blue population (blue line) becomes increasingly important
in the faintest magnitude bins.
As already mentioned in Sect.\,\ref{sect_intro}, many studies have
reported and described this colour bimodality in galaxies out to
$z\sim2$ \citep[e.g][]{Strateva2001,Blanton2003,Baldry2004,
Bell2004,Willmer2006,Faber2007,Blanton2007,Fritz2014}.

The optical colour distribution is in general well modeled by the sum of
two Gauss functions~\citep{Strateva2001,Baldry2004,Ball2008}.
Figure~\ref{gv_B_grid} also shows that the $UBV$ rest-frame colour
distribution is well approximated by the sum of two Gaussians (the brown
curves), in agreement with previous results~\citep[e.g.][]{Baldry2004}
Similar results are also found in e.g.~\citet{Ball2008}, \citet{Gonzalez2009}.

The mean and the dispersion of each Gaussian component (the blue and red
curves) depend on magnitude and redshift.
The blue objects are characterised by a larger dispersion in colour than
the red ones, which justifies the terms blue cloud and red sequence
generally used to characterize the two populations.

The local minimum is thought to be populated by objects that are
evolving from star forming to quiescent galaxies.
We did not find a significant excess of objects between the two main
galaxy populations with respect to the sum of the two Gauss functions,
meaning that there is no statistical evidence of a third population of
objects. This is at variance with the results of other analyses which
claim to find an excess of objects in the region between the two
peaks~\citep[e.g.][]{Wyder2007,Mendez2011,Schawinski2009,Coppa2011,Loh2010,Lackner2012,Brammer2009}.
This excess of galaxies is usually found in the distribution of
several colour indices, such as $U-B$~\citep{Nandra2007,Yan2011},
$U-V$~\citep{Brammer2009,Moresco2010} and
$NUV-r$~\citep{Wyder2007,Fritz2014}. In particular, \citet{Wyder2007}
show that the $NUV-r$ colour distribution is not strictly the sum of
two Gaussians, and \citet{Coppa2011}, using zCOSMOS data in the
redshift range $0.5<z<1.3$, reported a third galaxy population located
between the blue and red populations.
The lack of the third galaxy population located between two
Gaussians peaks is possibly related with the fine size of luminosity and
redshift bins used in this study.
The excess of galaxies with respect to the the sum of the two Gaussians
appears when using a coarser grid redshift or luminosity.
Moreover, the $UBV$ rest-frame colour, being an excellent proxy to sSFR,
is more efficient at separating different galaxy populations and less
prone contaminant objects that could populate the intermediate colours.

While in the local Universe the colour-magnitude diagram is
effective at dividing galaxies into different populations
\citep[e.g.][]{Strateva2001,Baldry2004,Wyder2007}, to study distant
galaxies it becomes important to consider how the selection depends
also on galaxy luminosity and redshift \citep{Bell2004}.  Exploring
the effects of the luminosity and redshifts in the VIPERS sample, we
reveal the systematic blueing of both the blue and red populations
moving towards fainter magnitudes at fixed redshift (blue and red
vertical lines in Fig.~\ref{gv_B_grid}). Quantitatively, the blue
cloud moves from $UBV=1.07$ to $0.73$ and the red sequence from
$UBV=1.50$ to $1.37$ at $z=[0.50,0.65]$ for values of $\Delta B_{\rm ev}$
increasing from $-1.5$ to $1.0$.  Similar trends in the
analysis of the $u-r$ rest-frame colour have been found in the low
redshift universe by \citet{Ball2008} and \citet{Mendez2011} using the
SDSS galaxy sample. Moreover, both populations in Fig.~\ref{gv_B_grid}
evolve toward bluer colours moving to higher redshifts.

The positions of the Gaussian maxima of the red and blue populations
can be described by the following formalism:
\begin{eqnarray}
UBV_{\rm b} &=& 1.06(\pm0.02)-0.36(\pm0.03)z-0.18(\pm0.01)\Delta B_{\rm ev} \label{colour_magnitude_b} \\
UBV_{\rm r} &=& 1.56(\pm0.02)-0.26(\pm0.02)z-0.06(\pm0.01)\Delta B_{\rm ev}
\label{colour_magnitude_r}
\end{eqnarray}
where $z$ is the redshift, $\Delta B_{\rm ev}$ is the distance from
the evolving characteristic luminosity as defined in
Eq.~\ref{Bstar_eq}, and $UBV_{\rm b}$ and $UBV_{\rm r}$ are the
central positions of the blue and red galaxy distributions,
respectively.  The quoted errors on the coefficients were estimated
through a bootstrap procedure using 1000 resamplings.


\section{S\'ersic index}
\label{sect_sersic_index}

\subsection{Estimation of S\'ersic parameters}
\label{sect_sersic}

To derive the surface brightness parameters of VIPERS galaxies, we
have performed a 2D fit of the observed galaxy $i$-band light
distribution with a PSF-convolved S\'{e}rsic model.  We used the
single component \citet{Sersic1963} profile given by the equation:
\begin{equation}
I(r)=I_e\exp\left\{ -b_n \left[ \left( \frac{r}{r_e}\right)^{1/n}-1\right] \right\},
\label{sersic_eq}
\end{equation}
where $r_e$ is the radius enclosing half of the total light of the
galaxy, $I_e$ is the mean surface brightness at $r_e$, and $b_n$ is a
normalization factor, which is chosen in such a way that $r_e$
corresponds to the half-light radius~\citep{Graham2005}.  This
parametrization well describes the light distributions of elliptical,
spiral and irregular galaxies \citep[see e.g.][]{Trujillo2001b}.  The
detailed analytical properties of Eq.~(\ref{sersic_eq}) are discussed
e.g.~by~\citet{Ciotti1999}, \citet{Trujillo2001}, \citet{Graham2005}.

To perform the fit we used the code GALFIT \citep{Peng2002}.  The
fitting procedure of GALFIT provides the value of the semi-major axis
($a_e$), the axial ratio ($b/a$) of the profile, from which the
circularised effective radius ($r_e=a_e\sqrt{b/a}$) is derived, as
well as the S\'{e}rsic index $n$, and the apparent magnitude of the
modeled galaxy.

\begin{figure}
\centering
\includegraphics[width=\hsize]{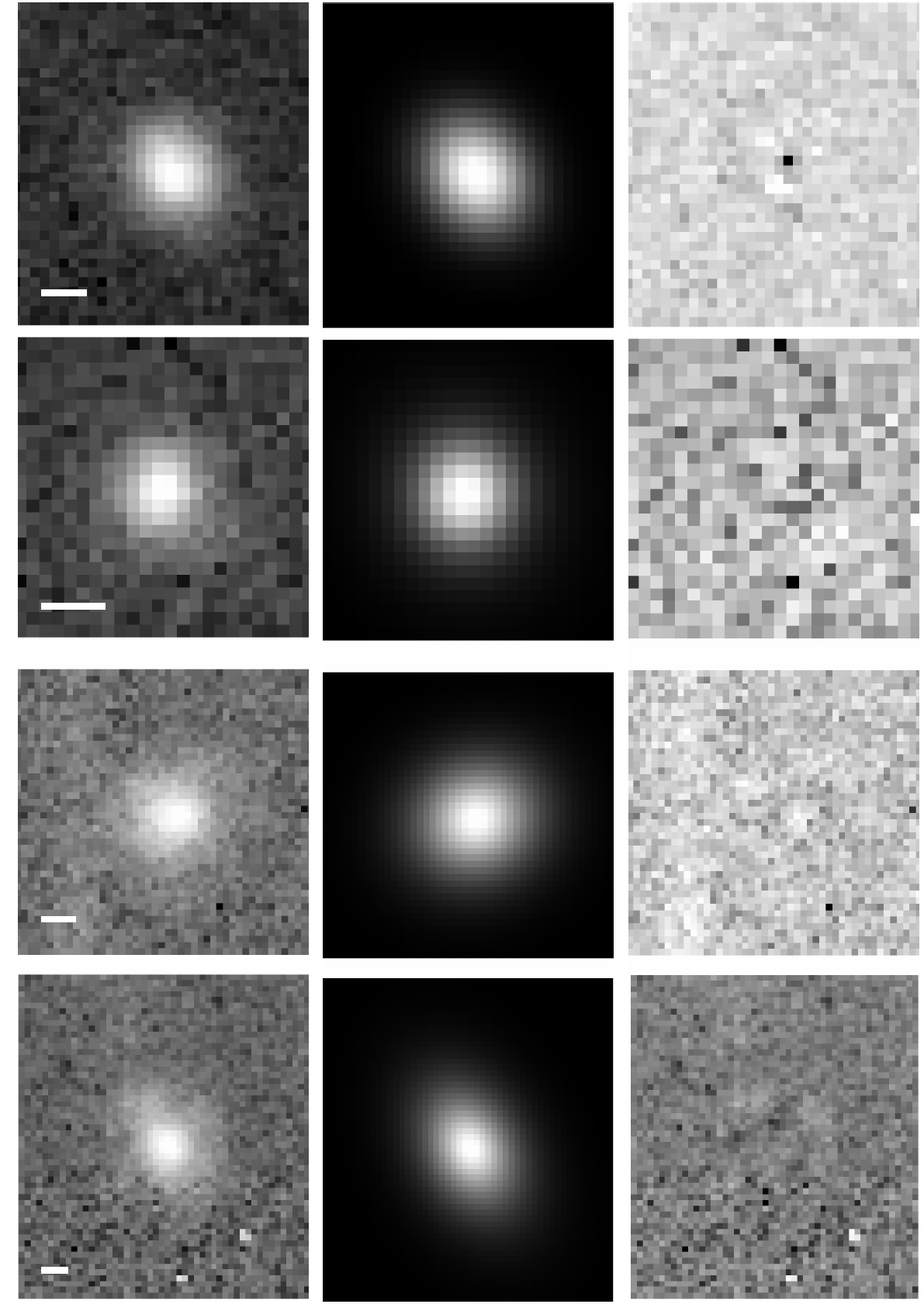}
\caption{Four examples of the GALFIT image decomposition procedure.
Left column: postage stamps with observed galaxies; middle column:
best-fit PSF-convolved S\'ersic model to each galaxy; right column:
residual images.
The small horizontal bars in the left column correspond to $1\arcsec$.}
\label{galfit_fit}
\end{figure}

We used the CFHTLS-T0006 images of VIPERS targets, and divided each
$1\degr\times1\degr$ tile into postage stamps centred on each VIPERS
galaxy (see some examples in Fig.~\ref{galfit_fit}).  To define the
size of the postage stamps, we rely on the \texttt{SExtractor}
parameters, which describe the ellipse associated to a given $i$-band
detection, namely $R_\mathrm{K}$ (KRON\_RADIUS), $A$, $B$ (A\_IMAGE,
B\_IMAGE) and $\theta$ (THETA\_IMAGE).  The centre of each postage
stamp coincides with the centroid of the \texttt{SExtractor} ellipse,
while its sides ($\Delta x$ and $\Delta y$) are four times larger than
the projected total dimension of the ellipse on the $x$ and $y$ axis,
i.e.
\begin{eqnarray}
\Delta x = 8R_\mathrm{K}\sqrt{(A\cos \theta)^2 + (B\sin \theta)^2}, \nonumber \\
\Delta y = 8R_\mathrm{K}\sqrt{(A\sin \theta)^2 + (B\cos \theta)^2}.
\label{postage_size_eq}
\end{eqnarray}
These sizes ensure that each postage stamp has sufficient object-free
pixels to estimate the background emission, which plays an important
role in galaxy image fitting.

There are two main approaches to estimating the level of background
emission.  In the first procedure the background is characterised
independently of the analysis of the target object, computed
\textit{a priori} e.g.~from an annular region surrounding the galaxy
\citep{Barden2005,Haussler2007,Guo2009,Fritz2009a,Fritz2009b}.  In the
second method the background is a free parameter that can vary during
the GALFIT fitting \citep{Mosleh2013,Cassata2011}.  The S\'ersic
parameters presented in this paper were obtained adopting this second
approach.  When the area of the postage stamp is $\approx10$ times
larger than the target galaxy and the sky-background variance is
uniform, the two methodologies to estimate the background are
equivalent \citep{Cassata2011,Mosleh2013}.  These conditions are
satisfied in our data.

Postage stamps centred on each galaxy were extracted from the CFHTLS
tiles, and SExtractor run to detect all of the objects contained
therein.  In the fitting procedure, all of the other objects within
the postage stamp are masked, unless the aperture ellipse of a
secondary object, increased by a factor $1.5$, overlaps with that of
the main target. In that case, the two (or more) photometric sources
are fitted simultaneously to get the best values of the S\'ersic
profile parameters.  The values MAG\_AUTO, FLUX\_RADIUS, A\_IMAGE,
B\_IMAGE, THETA\_IMAGE obtained by SExtractor were used as a first
guess of $r_e$, position angle, ellipticity, and magnitude in GALFIT.
In the absence of an estimate of the S\'ersic index $n$, the initial
value of this parameter in GALFIT fit was set to $n=1.7$ for all
galaxies.

GALFIT requires a local Point Spread Function (PSF) for each postage
stamp to convolve the S\'ersic model.  The PSF model is generated at
the centre of each galaxy using a 2D Chebyshev approximation of the
elliptical Moffat function parameters. A detailed description of the
PSF construction is given in Appendix~\ref{appendix_psf}.

Figure~\ref{galfit_fit} shows some examples of the fit performed by
GALFIT for VIPERS galaxies.  The original image of the galaxy, the
best-fit PSF-convolved S\'ersic model of each galaxy, and the residual
map (real image - model) are shown.

More details about our morphological analysis and the reliability of
GALFIT results are presented in Appendix~\ref{appendix_tests}.
Briefly, we added 4000 artificial galaxies to the CFHTLS images with
structural parameters generated from the S\'ersic indices, magnitudes
and effective radii obtained by GALFIT for a randomly-selected subset
of the VIPERS galaxies used in this analysis. From these tests we
estimate uncertainties in our measurements of $n$ of $|\Delta n|/n =
0.16$ at the 68\% level, and 0.33 at the 95\% level (i.e. for 95\% of
galaxies in our sample), while the effective radii are accurate to
within 4.4\% and 12\% for 68\% and 95\% of our sample respectively.
Our tests also confirm that any bias in the $n$ measurements is
negligible.

\begin{figure*}
\centering
\includegraphics[width=\hsize]{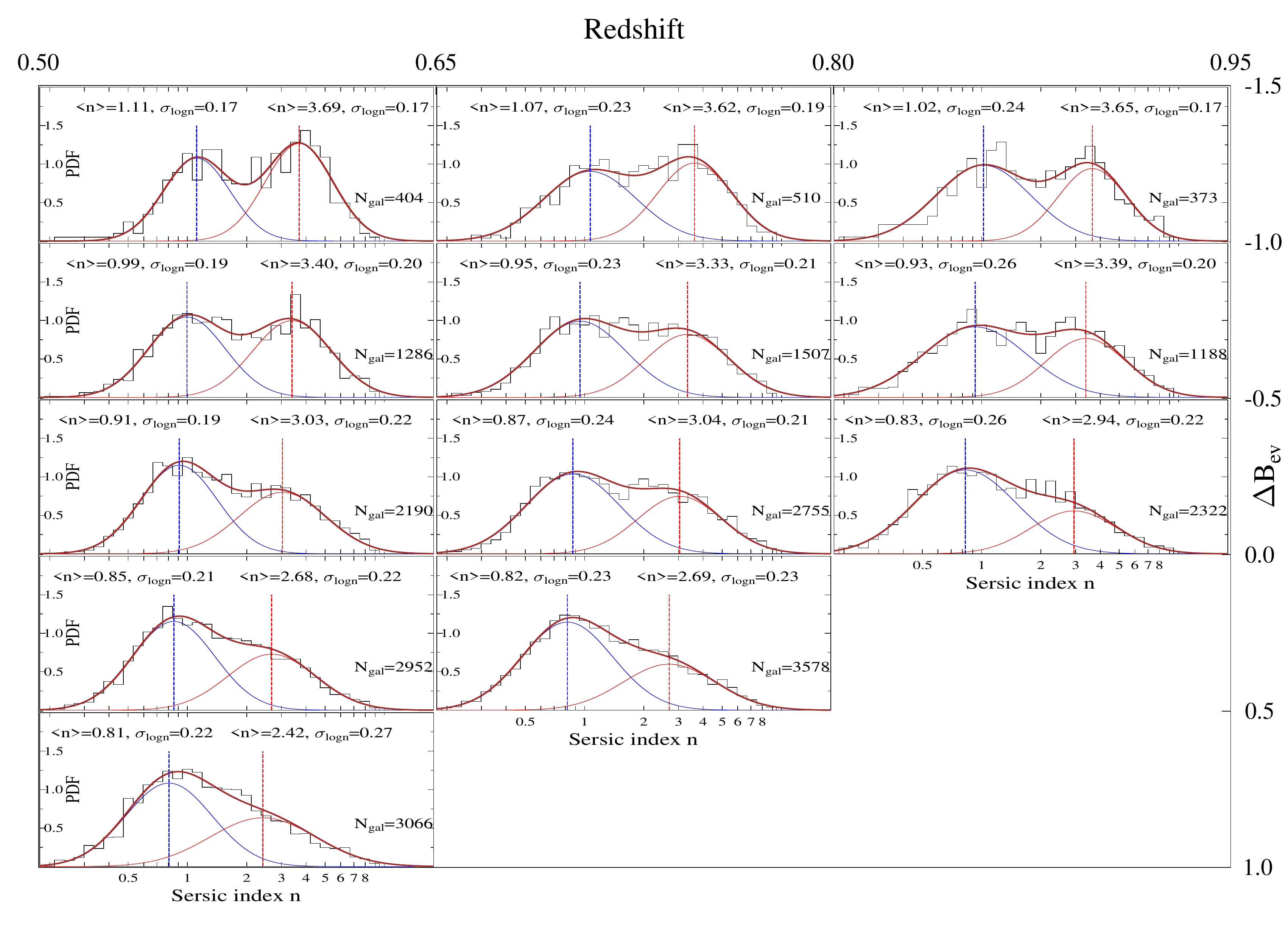}
\caption{The S\'ersic index distribution (black histograms) for
different redshift (from left to right) and $\Delta B_{\rm ev}$
luminosity bins (from top to bottom).
The blue and red solid lines show the Gaussian fits to the disc-like and
spheroid-like populations, respectively.
The vertical dashed lines mark the central values of each Gaussian.
The sum of the two Gaussian fits is shown as a solid brown line.
The central values $\langle n\rangle$ of the Gauss functions, their
$1\sigma$ widths, and the total number of galaxies in each bin are shown
in each panel.}
\label{grid_n}
\end{figure*}

To analyse the galaxies with the best quality S\'ersic function
parameters obtained from the data we selected only the objects with
reduced $\chi^2$ ($\chi^2_{DoF}$) values smaller than $1.2$, following
\citet{Peng2002}.
This removes only 4\% galaxies in the extreme high-end tail of the
$\chi^2_{DoF}$ distribution, with the vast majority of fits producing
$\chi^2_{DoF}$ values in the range 0.9--1.15.
We also discarded 261 objects with $n<0.2$: low values of the S\'ersic
index imply a lower accuracy in the approximation of the S\'ersic $b_n$
normalization factor \citep{Ciotti1999}
and introduces a small bias on the distribution of the disk-like
profiles, but it is negligible when compared to the error bars of the
fitted Sersic index.
Similar low-$n$ cuts are commonly used in other studies.

After removing 5707 (12\%) bad fitted by GALFIT objects, we
obtained our final sample, constituded by 38\,620 galaxies.
Finally, the volume-limited sample of objects presented in
Fig.~\ref{B-Bstar_all} consists of 22\,131 galaxies.

\subsection{S\'ersic index bimodality}
\label{sect_bimod_n}

Figure~\ref{grid_n} shows the S\'ersic index distribution of VIPERS
galaxies in the same luminosity and redshift bins as used in
Sect.~\ref{sect_bimod_c} for the $UBV$ colour. Since the S\'ersic
index, $n$, appears as an exponent in Eq.~\ref{sersic_eq} defining the
S\'ersic profile~\citep{Driver2006,Driver2011}, a logarithmic-spaced
$x$-axis is used to optimise the analysis and visualisation of the
wide range of $n$ values.

Similarly to the $UBV$ histograms shown in Fig.\,3, the S\'ersic index
distribution is bimodal in many of the redshift--luminosity bins.
We thus fit each S\'ersic index distribution as a sum of two Gauss
functions in $\log n$, with one Gaussian component considered to
represent the disk-like population (blue curves), and a second to
represent the spheroid-like galaxy population (red curves).
The sum of the two Gaussian fits (solid brown curves) describes well the
S\'ersic index distribution at all redshifts and luminosities explored
here.
Even though for galaxies fainter than the characteristic luminosity of
the LF, i.e. $\Delta B_{\rm ev}>0.0$, the global distribution is not so
evidently bimodal, nonetheless it is well reproduced by the sum of the
two Gauss functions.

The vertical blue and red dashed lines in Fig.~\ref{grid_n} show the
central values of the two Gaussian components for each redshift and
luminosity bin. Comparing the locations of these lines from panel to
panel, we see that the mean S\'ersic indices of both disk-like and
spheroid-like galaxy populations vary systematically with luminosity
and redshift. In particular, both disk-like and spheroid-like
populations become increasingly concentrated with increasing
luminosity and decreasing redshift.

We find that the best two-dimensional linear fit of these positions in
the redshift $z$ versus $\Delta B_{\rm ev}$ luminosity plane is well
described by the following equations:
\begin{eqnarray}
\log n_{\rm d} &=& 0.04(\pm0.01)-0.16(\pm0.01)z-0.07(\pm0.01)\Delta B_{\rm ev} \label{disk_one_dim} \\
\log n_{\rm s} &=& 0.47(\pm0.01)-0.03(\pm0.01)z-0.09(\pm0.01)\Delta B_{\rm ev} \label{sphr_one_dim}
\end{eqnarray}
where $\Delta B_{\rm ev}$ is the luminosity given by
Eq.~\ref{Bstar_eq} and $n_{\rm d}$ and $n_{\rm s}$ are the mean
S\'ersic indices of the disc-like and spheroid-like galaxy
populations.  The errors of the best-fit coefficients were estimated
by a bootstrap procedure using $1000$ resamplings.

The S\'ersic index $n = 1$, commonly used to model the light profile of
the disk-like galaxies, is well in the range [0.81,1.11] spanned by the
average S\'ersic indexes measured within the analysed
redshift-luminosity space limits.
For spheroid-like galaxies we find mean values in the range
$[2.42,3.69]$, lower than the typical value used to describe
nearby elliptical galaxies \citep[i.e., $n=4$, see][]{deVaucouleurs1948}.
Other authors have reported similar S\'ersic indices for
early-type galaxies, e.g. $\langle n\rangle=3.0$ \citep{DOnofrio2001},
$\langle n\rangle=3.3$ \citep{Padmanabhan2004}, $n>2.5$
\citep{Eales2015,Griffith2012}.
Moreover, the tests presented in the Appendix ensure that the S\'ersic
parameters we have obtained are reliable and that the bias in the
estimate of $n$ is negligible for all the redshift and luminosity bins
considered in this analysis.


\section{Comparison with literature}
\label{sect_literat}

Previous studies have shown that the S\'ersic index of galaxies depends
on both their absolute magnitude and redshift~\citep[e.g][]{Graham2003,  
Tamm2006,vanDokkum2010, vanDokkum2013,Patel2013,Buitrago2013}. To
compare our results with other works Eqs.~\ref{disk_one_dim} and
\ref{sphr_one_dim} are combined with Eq.\ref{Bstar_eq} to obtain the
following relations:
\begin{eqnarray}
\log n_{\rm d} &=& -(M_{B}+19.30+3.74z)/13.75 \label{disk_Bz_mag} \\
\log n_{\rm s} &=& -(M_{B}+14.87+1.87z)/10.68 \label{sphr_Bz_mag}
\end{eqnarray}
where the dependence on absolute magnitude is made explicit.
The relation between galaxy luminosity and S\'ersic index has been
reported in many studies for spheroid-like galaxies \citep[e.g.
][]{Young1994,Graham2003,Ferrarese2006}.
Moreover, a link between structural parameters and luminosity has
also been studied by~\citet{Cross2004} for E/S0 galaxies in the redshift
range from $z=0.5$ to $1$.
Equations~\ref{disk_Bz_mag} and \ref{sphr_Bz_mag} show that fainter disc
and spheroidal galaxies have lower value of the S\'ersic index than the
luminous ones and that this relation depends on redshift.

\begin{figure}
\centering
\includegraphics[width=\hsize]{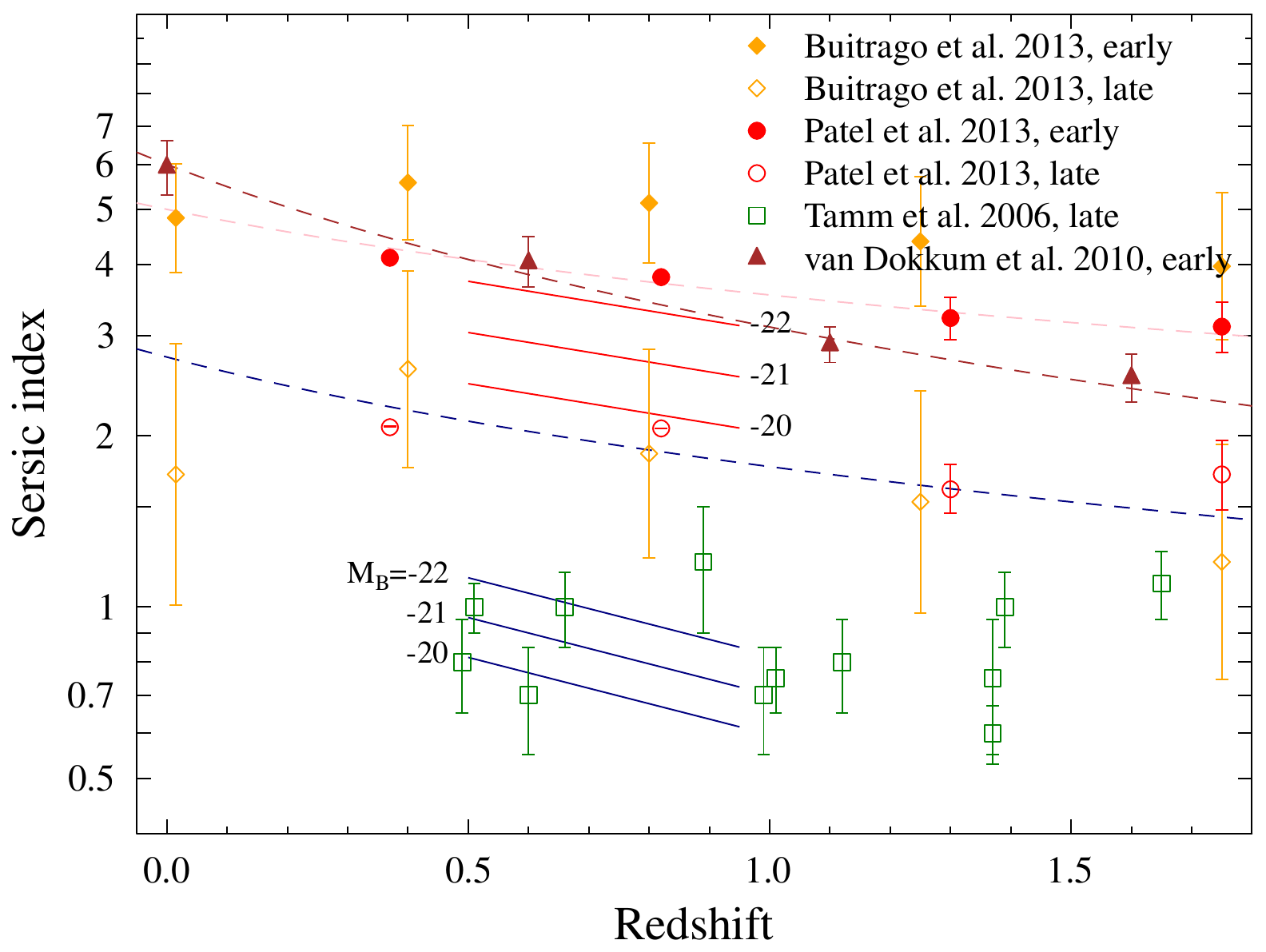}
\caption{S\'ersic index -- redshift relation: VIPERS results are
presented as red and blue solid lines for spheroid- and disc-like
galaxies respectively for three values of $B$-band absolute magnitude.
\citet{Patel2013}'s results for quiescent and star-forming objects with
$\log(M/M_\odot)>10.5$ are shown as red filled and empty circles and
dashed orange and blue lines.
The relation found by \citet{vanDokkum2010} for a constant comoving
number density sample is plotted in brown (short-dashed line and
triangles).
\citet{Buitrago2013}'s results for a sample visually classified into
early- and late-type galaxies is shown as orange filled and empty
diamonds, while disc-galaxies measured by \citet{Tamm2006} are
represented by green squares.}
\label{redshift_n}
\end{figure}

The dependence of S\'ersic index on redshift has been analysed in many
studies
\citep[e.g.][]{Tamm2006,vanDokkum2010,vanDokkum2013,Patel2013,Buitrago2013}.
Figure~\ref{redshift_n} shows the S\'ersic index-redshift relations for both
disc-like and spheroid-like populations within VIPERS for three absolute
magnitude values and the comparisons with previous studies. In the next
sections we analyse in detail the comparison for the two classes of
galaxies.

\subsection{Spheroid-like galaxies}

\citet{Patel2013} computed structural parameters of massive galaxies
in high-resolution HST imaging from the CANDELS and COSMOS surveys, and
measured the evolution of the S\'ersic index of galaxies in the redshift
range $0.25<z<3$, after splitting them into quiescent and star-forming
populations on the basis of their rest-frame $UVJ$ colours.

The solid red lines show the S\'ersic index-redshift relations for
spheroid-like populations described by Eq.~\ref{sphr_Bz_mag} for three
values of $B$-band absolute magnitude.
The solid red circles in Fig.~\ref{redshift_n} show the median S\'ersic
indices for quiescent galaxies with $\log(M/M_\odot)>10.5$ in four
redshift bins, while the orange dashed line indicates their best-fit
S\'ersic index--redshift relation over the redshift range $0<z<2.5$ of
the form $n \propto (1+z)^{-0.50(\pm0.18)}$.
The exponent of this relation is consistent with our fit,
$n\propto(1+z)^{-0.64}$, we obtain for our brighest ($M_{B}=-22$)
spheroid-like galaxies, which also fulfill their criterion
$\log(M/M_\odot)>10.5$.

\citet{vanDokkum2010} measured the S\'ersic index parameter from
stacked rest-frame $R$-band (observed $J,H$-band) images from NEWFIRM
Medium Band Survey. They selected a sample at a given constant
cumulative number density, which results in their use of a stellar mass
limit which evolves with redshift.
The stellar mass limit of their selection at our mean redshift $z\sim
0.7$ is $\log(M/M_\odot)>11.35$ and does not vary considerably
($<0.07$\,dex) in the redshift range we are exploring, $0.5<z<0.95$.
At these large stellar masses the galaxy population is dominated by
quiescent objects.
Rather than fitting S\'ersic profiles to each individual galaxy, and
measuring the mean of the distribution, \citet{vanDokkum2010} created
deconvolved stacked images of massive galaxies within bins of redshift
and fitted S\'ersic functions to the stacked radial surface density
profile, the results of which are shown as brown triangles in
Fig.~\ref{redshift_n}.
They measure a best-fit evolution for the S\'ersic index of the form
$n=6.0\times(1+z)^{-0.95}$ over the range $0<z<2$
0<z<2, presented with the brown line in this plot.
The redshift evolution is faster than in~\cite{Patel2013}, perhaps
reflecting the contamination by non-quiescent objects or systematics in
measuring S\'ersic indices from stacked images, but in good agreement
with our results in the common $z$-range.

\citet{Buitrago2013} estimated quantitative and visual morphologies
from HST images of a sample drawn from the DEEP2 and GOODS surveys,
combined with a local sample based on SDSS imaging.
Their sample of massive $\log(M/M_\odot)>11$ galaxies was then
subdivided into early- and late-type galaxies on the basis of the visual
classification.
The mean S\'ersic indices of visually-selected early-types in bins
of redshift are displayed as magenta crosses, and show the same
gradual increase in $n$ with time, albeit systematically shifted to
higher S\'ersic index values by $\Delta n\sim1.5$.
Despite of the different selection criteria, the evolution of the S\'ersic
index for bright spheroid-like is in good agreement with the relations
found in literature for massive quiescent galaxies.

\subsection{Disk-like galaxies}

The S\'ersic index -- redshift relation for disc-like galaxies given by
Eq.~\ref{disk_Bz_mag} is presented in Fig.~\ref{redshift_n} by the dark
blue lines, and for $M_{B}=-22$\,mag can be written as $n_{\rm d} =
1.65(1+z)^{-0.98}$.
The dependence of the $n$-redshift relation on absolute magnitude is
smaller for disc-like galaxies than for spheroid-like ones, while its
evolution with cosmic time is faster for disc-like galaxies than for
spheroid-like ones.

\citet{Patel2013} and \citet{Buitrago2013} found similar, decreasing
trends.
However, their relations are significantly offset from our results by
$\Delta n\sim1$, probably reflecting the fact that they used selection
criteria very different from ours (i.e. star-forming galaxies in
\citealt{Patel2013} and very massive visually classified late-type
galaxies in~\citealt{Buitrago2013}).
In particular, we found that the characteristic stellar mass, estimated
from the  mass -- luminosity relation, of our disc-like sample
corresponds to a selection of stellar masses smaller than
\mbox{$\log(M/M_\odot)=10.5$}.

For a much more meaningful comparison we turned to the~\cite{Tamm2006}
sample who measured the S\'ersic profile of 22 galaxies in the HDF-S
using a selection similar to ours, as they have only considered
disk-like galaxies (with $n<2$) in absolute magnitude range
$-17<M_B<-22$. It is therefore reassuring that theirs results are
consistent with ours, as shown in Fig.~\ref{redshift_n}.

Comparing our results with previous work, we find a general good
agreement of the evolution of the S\'ersic index for spheroid-like galaxies
with the ones for quiescent and early-type galaxies.
Instead, galaxies defined as star-forming are characterised by larger
values of S\'ersic index when compared to disk-like ones.


\bigskip
\section{S\'ersic index -- colour distribution}
\label{sect_n_col}

In Sects.~\ref{sect_bimod_c} and \ref{sect_bimod_n} we independently
analysed the $UBV$ rest-frame colour and the logarithm of the S\'ersic
index $n$ of the VIPERS galaxies as a function of the redshift $z$ and
$\Delta B_{\rm ev}$ luminosity. Both parameters show a bimodal
distribution. Using the local galaxy sample of the Millennium Galaxy
Catalogue, \citet{Driver2006} showed not only that both colour and
S\'ersic index are characterised by bimodal distributions, but that
two well-separated populations exist on the $u-r$ rest-frame colour
versus $\log(n)$ plane.

\begin{figure*}
\centering
\includegraphics[width=0.81\hsize]{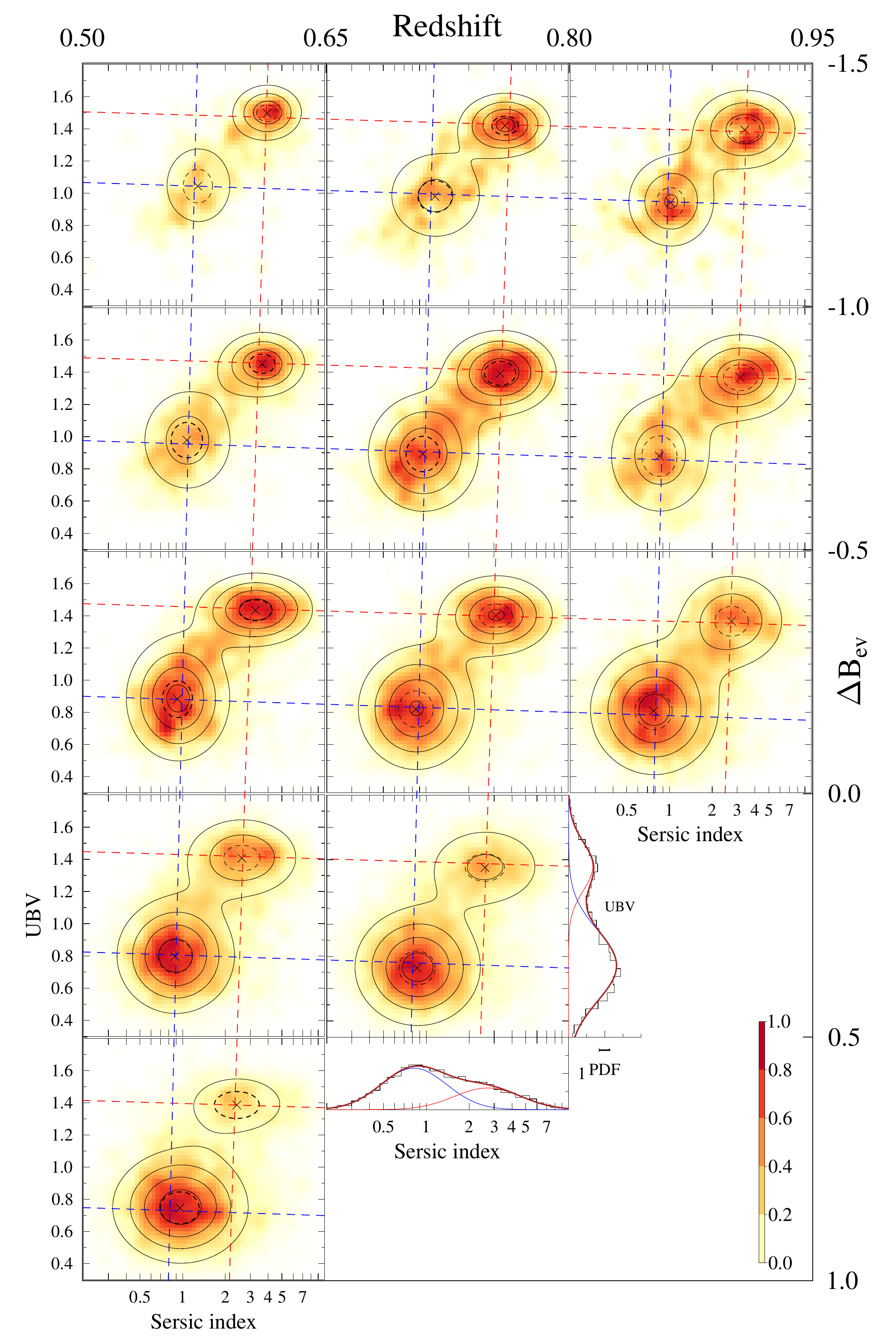}
\caption{$UBV$ rest-frame colour vs S\'ersic index $n$ distribution.
Each panel shows the colour coded galaxy surface density distribution
map of the VIPERS galaxies in each redshift and $\Delta B_{\rm ev}$
luminosity bin.
The colour bar presented in the bottom right corner gives the normalised
galaxy surface density.
The contour lines show in steps of 0.2 the density values obtained from
the two Gaussians bivariate fitting procedure.
The blue and red dashed lines show the 2-dimensional model whereas the
crosses identify the positions of the centre of the given galaxy
population.
The dashed lines around the peaks shows the value of $0.5\sigma$ of the
Gaussian fit.
The histograms present the galaxy distributions projected on the
S\'ersic index and $UBV$ colour axes, at the bin $z=[0.65, 0.80]$ and
$\Delta B_{\rm ev}=[0.5, 0.0]$.
}
\label{ubv_n_dens_grid}
\end{figure*}

To investigate whether this is still true at high redshift we have
repeated \citet{Driver2006} analysis in each of our subsamples.
The results are shown in Fig.~\ref{ubv_n_dens_grid}.
The colours in each surface density map of this plot are normalised to
have values in the range 0--1, so that $1$ (dark red colour) is the peak
density in each bin.
The joint probability distribution of $UBV$ rest-frame colour and
S\'ersic index $n$ is clearly bimodal in all panels, with two well
separated peaks and indicates the presence of two different populations
that we identify with early- and late-type galaxies.

The plot shows that the distribution of the late-type galaxies are
centred at the S\'ersic index value $n \approx 1$ and the rest-frame
colour $UBV\approx 0.8$, while those of the early-type galaxies are
centred at $UBV \sim 1.4$ and $2.5<n<4$.
The latter peak appears somewhat elongated along the $n$-axis and moves
towards larger values of the S\'ersic index (from $n \sim 2.5$ to $4$)
with cosmic time, i.e. galaxies become more concentrated at lower
redshift.
The two peaks are separated by the local minimum located at
$UBV\sim1.2$, corresponding to sSFR $10^{-10}{\rm yr}^{-1}$ (see
Fig.~\ref{BV_UB}); a value that is often used to separate active from
passive objects~\citep[e.g.][]{Davidzon2016}.
From these plots we see that a more effective separation can be done
using the combined S\'ersic index $n$ and $UBV$ rest-frame colour
information.

We fitted the joint probability distribution of S\'ersic index and
$UBV$ colour in each redshift-luminosity bin with the sum of two
2D-Gaussians.
The iso-density contour lines are separated in step of 0.2 the
surface distribution density.
The dashed circle around each peak shows the $0.5\sigma$ level of each
2D-Gaussian.

We do not include a covariance term for the Gauss functions, to avoid
artificially creating apparent correlations between $UBV$ and $n$ within
the single populations due to the presence of the second population.

In addition to the dominant populations of early- and late-type
galaxies, Fig.\ref{ubv_n_dens_grid} shows that a fraction of blue
galaxies have large values of the S\'ersic index ($n\ga 2$), while
conversely some red galaxies have a S\'ersic index $n \approx 1$,
typical of disc-like objects.
We postpone a thorough investigation of these peculiar objects to a
future analysis.

Moreover, Fig.~\ref{ubv_n_dens_grid} gives us information on the
galaxy morphological type fraction in each luminosity/redshift bin.
It shows that the most luminous bins are dominated by early-type
galaxies, whereas the late-like galaxies dominate the less luminous
subsamples.

\begin{figure*}
\sidecaption
\includegraphics[width=12cm]{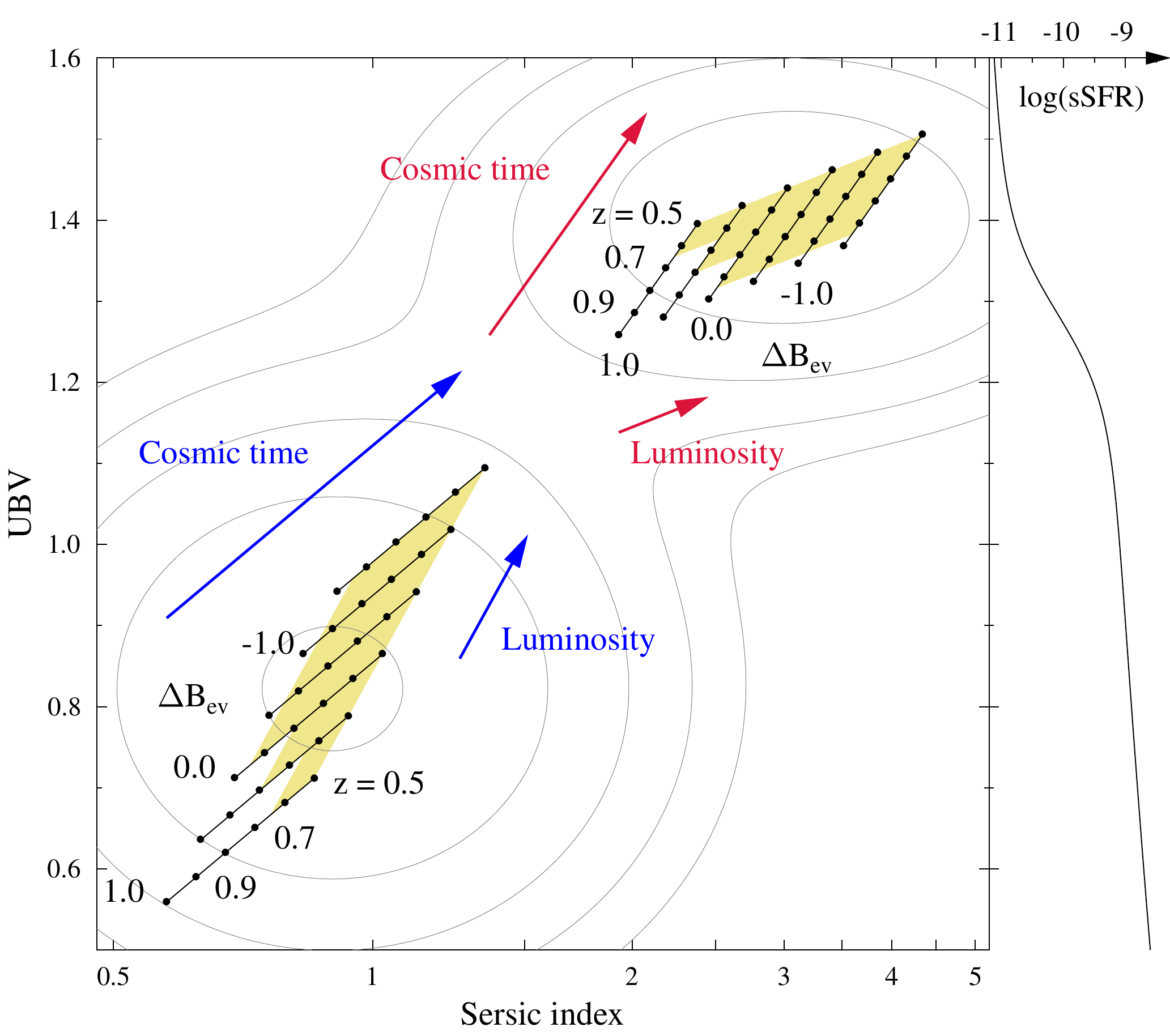}
\caption{$UBV$ rest-frame colour versus S\'ersic index $\log(n)$
relation of late-type (lower left corner) and early-type (upper right
corner) galaxies.
Dots indicate redshift from $z=0.5$ to $1.0$ in steps of $0.1$.
Black solid lines connect the values of $\Delta B_{\rm ev}$ from $-1.5$
to $1$ in $0.5$ magnitude steps.
The arrows show the direction of the redshift and luminosity galaxy evolution.
The right plot presents the $UBV$ colour versus sSFR relation.
The $0.2$\,dex background contour lines show the bivariate number
density of all studied VIPERS galaxies in our sample ($0.5<z<0.95$).
The yellow coloured regions mark the analysed redshift and luminosity
limits, as presented in Fig.~\ref{B-Bstar_all} and \ref{ubv_n_dens_grid}.
}
\label{n_ubv_model}
\end{figure*}

\subsection{Early-type galaxies in the $n$--$UBV$ plane}

The galaxy surface distributions presented in Fig.~\ref{ubv_n_dens_grid}
show that the positions of early- and late-type galaxy populations
change with both redshift and luminosity.

Fitting the sum of two 2D Gauss functions to the distributions in each
bin we obtain the positions of the population centres ($\log(n)$,
$UBV$).
Using these positions we determined the empirical relation connecting the
galaxy population centre with $\Delta B_{\rm ev}$ and redshift.
Our results in Sects.\ref{sect_bimod_c} and \ref{sect_bimod_n} showed
that the $UBV$ rest-frame colour and the S\'ersic index $\log(n)$
are well reproduced by a linear dependence on redshift and luminosity.
We thus fit the position of the early-type galaxy population centre
$(\log(n_{\rm e}), UBV_{\rm e})$ in Fig.~\ref{ubv_n_dens_grid} with a
two-dimensional linear function, obtaining the following set of
equations describing the central position of this galaxy population as a
function of redshift and luminosity:
\begin{eqnarray}
UBV_{\rm e}     = 1.58(\pm0.02)-0.27(\pm0.03)z-0.04(\pm0.01)\Delta B_{\rm ev}\ \label{early_model_ubv}\\
\log(n_{\rm e}) = 0.57(\pm0.03)-0.18(\pm0.04)z-0.10(\pm0.01)\Delta B_{\rm ev}, \label{early_model_n}
\end{eqnarray}
where $\Delta B_{\rm ev}$ is given by Eq.~\ref{Bstar_eq}.  The errors
of the fitted coefficients were estimated by a bootstrap procedure
using $1000$ resamples.

The relations given by Eqs.~\ref{early_model_ubv} and
\ref{early_model_n} were used to compute the central $UBV$ and $n$
values for the early-type galaxy populations in each redshift and
luminosity bin and are shown as crosses in Fig.~\ref{ubv_n_dens_grid}.
Comparing these positions with the shape of the higher density contour
lines and the $0.5\sigma$ widths of the 2D Gaussian fits (marked as
dashed ellipses) we find that the simple linear approximation given
above well predicts the observed peak position of the early-type galaxy
population.
The mean distance between the maxima positions from data and linear
model is smaller than $0.1\sigma$.

\subsection{Late-type galaxies in the $n$--$UBV$ plane}

The same procedure was also applied to the late-type galaxy
distributions.  The following set of equations describes the central
position $(UBV_{\rm l}$, $\log(n_{\rm l})$ of the late-type galaxy
population as a function of redshift and luminosity:
\begin{eqnarray}
UBV_{\rm l}    =  1.02(\pm0.03)-0.31(\pm0.05)z-0.15(\pm0.01)\Delta B_{\rm ev}\  \label{late_model_ubv}\\
\log(n_{\rm l})=  0.18(\pm0.03)-0.34(\pm0.04)z-0.08(\pm0.01)\Delta B_{\rm ev}. \label{late_model_n} 
\end{eqnarray}
The crosses in Fig.~\ref{ubv_n_dens_grid} show the position of the
late-type galaxies distribution centre given by
Eqs.~\ref{late_model_ubv} and \ref{late_model_n}.
The plot shows that our linear model well reproduces the positions of
galaxy density maxima, with distance from the maxima computed from data
smaller than $0.1\sigma$.

\subsection{Comparison of 1D to 2D approximation}

In Sects.\ref{sect_bimod_c} and \ref{sect_bimod_n} we focussed our
attention on the 1D distributions of the S\'ersic index and $UBV$
rest-frame colour.  It is worth comparing those results with the ones
obtained from the 2D approximation.  We find that both approaches give
almost the same results for the $UBV$ rest-frame colour position of
the galaxy populations centres.  The 1D relations given by
Eqs.~\ref{colour_magnitude_b} and \ref{colour_magnitude_r} and the 2D
ones presented by Eqs.~\ref{early_model_ubv}, \ref{late_model_ubv} are
consistent with each other within $\pm 1\sigma$ of the fitted
parameters.

However, some significant differences occur in the approximation of
the S\'ersic index $\log(n)$ positions.  The coefficients representing
the redshift dependence in the 1D relations given by
Eqs.~\ref{disk_one_dim}, \ref{sphr_one_dim} and the 2D ones presented
by Eqs.~\ref{early_model_n}, \ref{late_model_n} are different, with
the redshift dependence in the 1D representation being significantly
shallower than the one obtained with the 2D analysis.  The origin of
this difference is evident when comparing Figs.~\ref{ubv_n_dens_grid}
and Fig.~\ref{grid_n}: the 2D galaxy distribution very well separates
both galaxy populations for all $\Delta B_{\rm ev}$ luminosity and
redshift bins. In contrast, in the 1D projection of the S\'ersic index
$\log(n)$ these distributions partially overlap each other, especially
for the less luminous disc- and spheroid-like galaxy populations, as
clearly seen in the histograms presented in Fig.~\ref{grid_n}.
Because of this, the results obtained with the 2D approach are much
better determined and more robust than those obtained with the 1D
analysis.



\section{Sersic index -- $UBV$ colour coevolution}
\label{sect_n_col_evol}

The analysis presented in the previous sections provides a
quantitative description of the S\'ersic index -- $UBV$ colour relation
and its dependence on redshift and the galaxy luminosity.
Figure~\ref{n_ubv_model} makes use of Eqs.~\ref{early_model_ubv},
\ref{early_model_n}, \ref{late_model_ubv} and \ref{late_model_n} to
present these dependencies on the S\'ersic index versus $UBV$ colour plane.
Dots represent values given by the equations presented in the
previous sections, for redshift from $z=0.5$ to 1.0 in steps of
0.1, and black lines connect points corresponding to the fixed values of
$\Delta B_{\rm ev}$ ranging from -1.5 to 1.0 in steps of 0.5\,mag.
Contour lines represent the galaxy surface density of the whole VIPERS galaxy sample
studied in this paper, in steps of $0.2$ dex. The coloured
regions highlight the redshift and luminosity limits presented in
Figs.~\ref{B-Bstar_all} and~\ref{ubv_n_dens_grid}.
The blue and red arrows indicate the change of values of $UBV$ and S\'ersic
index $\log(n)$ as a function of redshift and luminosity.
In Sec.~3.1 we show that the $UBV$ rest-frame colour is well
correlated with the sSFR.
The approximate relation of $UBV$ colour versus sSFR is presented on the
right side of Fig.~\ref{n_ubv_model}.

The $UBV$ rest-frame colour versus $\log(n)$ diagram allows to make a
division, at the intermediate redshift $z\approx0.7$, between the
late-type galaxies (presumably, disk-like and blue, mostly star-forming)
with $UBV<1.2$ and $n<1.5$ and early-type galaxies (presumably,
spheroidal and red, mostly quiescent) for which $UBV>1.2$ and
$n>1.5$.

Figure~\ref{n_ubv_model} visually connects four galaxy parameters and
allows to present the coevolution of the properties of galaxies
belonging to the early- and late-type classes.
In fact, from this figure it is already clear that the evolution of the
relation between $UBV$ and $n$ is markedly different for early- and
late-type galaxies,  like also other studies
found~\citep[e.g.][]{Blanton2003}.
We also find that the S\'ersic index $n$ of both main morphological 
galaxy types - disk-like and spheroidal - increases both with their
luminosity  and cosmic time.
This result is consistent with observations and numerical
simulations~\citep[e.g.][]{Conselice2003,Conselice2005,
Treu2005,Bundy2005,Brook2006,Aceves2006,Hopkins2007}.

\subsection{Early-type galaxies}

The results presented in the previous sections allow us to give a
general overview of the colours and structural properties of
early-type galaxies (ETGs).  Figure~\ref{n_ubv_model} shows explicitly
the effect of evolution and luminosity on the colours and
structural properties of ETGs.
Firstly, it confirms that ETGs simultaneously become redder and more
concentrated {\em both} with cosmic time and increasing luminosity
(presumably correlated with stellar
mass)~\citep[][]{Trujillo2001,Graham2003,Tamm2006,
vanDokkum2010,vanDokkum2013,Patel2013,Buitrago2013}.
However, the effects of increasing luminosity and cosmic time on
early-type galaxies act in different directions.  This means that we
cannot take a low-luminosity early-type galaxy at $z=1.0$, and simply
wait a few Gyr for it to become as red {\em and} as concentrated as
its high-luminosity counterpart was at $z=1.0$.
At $z=1.0$, we see that a low-luminosity ($\Delta B_{\rm ev}=+1.0$) red
galaxy is 0.10\,mag bluer in $UBV$ and  0.6 times less concentrated than
its 10 times more luminous ($\Delta B_{\rm ev}=-1.5$) red counterpart.

Following a galaxy evolutionary track, we see that while a galaxy can
rapidly redden to match its high-luminosity counterpart by $z=0.63$,
over the same time-scale it only marginally increases its
concentration by a factor equal to 1.17, i.e. only a quarter of the
amount needed to match that of high-luminosity ETGs at $z=1.0$.
Indeed, even at $z=0$ (assuming an extrapolation of the linear trends)
its S\'ersic index will not have increased sufficiently.

Low-luminosity early-types are known to have later formation epochs and
more extended bursts of star formation than their high-luminosity
counterparts and have delayed star formation histories (e.g. Thomas et
al. 2005).
The delayed star formation can be also seen tentatively from the plot in
Fig.~\ref{n_ubv_model}, where low luminosity galaxies seem to have on
average larger values of sSFR than brighter ones at a fixed redshift.

The results presented here confirm that while it is possible to account
for this delay by matching low-luminosity ETGs observed at lower
redshifts to higher mass ETGs seen at earlier epochs, and to first order
to have stellar populations of equivalent ages (although the
metallicities will differ), the lower-luminosity ETGs will still have
quite different structural properties, being much less concentrated at
fixed stellar age.
This fundamental difference likely reflects the less active merger
history of lower luminosity (mass) ETGs~\citep[e.g.][]{Rodriguez2016,
Lacey1993, Aceves2006, DeLucia2006}, meaning they cannot build up the
more extended stellar halos of high-mass ETGs.

If we assume that the increase in $n$ is due to major
mergers~\citep[e.g][]{Aceves2006} and the continual accretion of
material onto the outskirts of the galaxy, the trends of
Figure~\ref{n_ubv_model} suggest that low-luminosity ETGs do not undergo
sufficient minor mergers at late epochs to ``catch up'' the much more
active merger history of high-mass ETGs at $z>1$.

\subsection{Late-type galaxies}

At first sight, Figure~\ref{n_ubv_model} suggests that late-type
galaxies (LTGs) show very similar trends to early-type, becoming
simultaneously redder and more concentrated, both with cosmic time
(decreasing $z$) and increasing luminosity.
Moreover, the evolution from $z=1$ to $z=0.5$ in the $UBV$ vs. $\log(n)$
plane is similar in magnitude and direction to that of the early-type
population, leaving the separation between the two populations virtually
unchanged, as is presented in Fig.~\ref{ubv_n_dens_grid}.
Hence, the bimodality appears to neither strengthen or weaken with time,
at least for the redshift range studied here.

Interestingly however, the relative impacts of time and luminosity on
$UBV$ colour and $\log(n)$ appear to have flipped in comparison to that
seen among the ETGs.
The concentration of LTGs is most dependent on cosmic time, while $UBV$
colour increases mostly with luminosity.
At $z=1.0$, a low-luminosity LTG ($\Delta B_{\rm ev}=+1.0$) is
0.375\,mag bluer and 1.6 times less concentrated than its 10
times more luminous counterpart ($\Delta B_{\rm ev}=-1.5$).
By following its evolutionary track, it is able to change its structure
sufficiently rapidly to match the S\'ersic index of its high-luminosity
counterpart by $z=0.41$, but over this same time period it is only
expected to become 0.18\,mag redder, half of that required to match the
$UBV$ colour of the high-luminosity LTG at $z=1$.

Given the well known systematic decline in specific-SFRs among LTGs over
$0<z<1$, in which both high- and low-luminosity (stellar mass) spirals see their
star formation drop exponentially and in step~\citep[e.g.][]{Noeske2007,
Zheng2007}, it is intersting to note that their structural parameters
are changing more rapidly than their colours, while $UBV$ colour is more
dependent on luminosity.
It should in fact be easier to make a spiral galaxy redder by reducing
star formation, than an early-type galaxy, as the response to a
reduction in star formation is greatest when the galaxy is initially
blue (see e.g. Fig.~\ref{BV_UB} and the right plot of
Fig.~\ref{n_ubv_model}).
One explanation could be that the large change in $UBV$ colour with
luminosity among LTGs reflects more the increased reddening due to dust
in massive spirals rather than a decrease in specific-SFR.

Theoretically, it is expected that the bulge fraction of merger remnants
increases with the decreasing gas fraction of the
progenitors~\citep[e.g.][]{Robertson2006, Hopkins2009}.
The higher luminosity (mass) disk-like galaxies have a higher bulge
fraction due to major- and intermediate-mass ratio mergers.
The dense luminous part of galaxies are undisturbed during this process
and luminous material dominates the central regions of mergers' remnants
\citep{Barnes1992a} and their S\'ersic index value increases, as is
shown in this study.



\section{Summary}
\label{sect_concl}

In this paper we presented the coevolution of the galaxy morphological
properties and colours over the redshift range from $z=0.5$ to 1
combining high-quality imaging data from the CFHT Legacy Survey with the
large number of redshift and extended photometry from the VIPERS survey.
We used this new dataset to investigate the coevolution of galaxy
S\'ersic index and UBV rest-frame colour.
The galaxy structural parameters were measured by GALFIT fitting
the S\'ersic profile to the $i-$band CFHTLS T0006 images.
To do this, the PSF of the images was precisely estimated and
approximated over the whole of each $1\degr\times1\degr$ tile.
The resultant parameters were carefully tested by a set of different
methods which confirms the good quality of the fits and reliability of
their fitted values.
Our results can be summarized as follows:

\begin{itemize}
\item We find a clear bimodality of the $UBV$ rest-frame colour and
S\'ersic index distribution, very well approximated by a sum of two
Gaussians over the explored redshift and luminosity ranges.
We parametrised the position of the two maxima in $UBV$ and $n$
distributions as a function of luminosity and redshift.
This parametrization allow us to analyse the colours and structural
parameters of the red and blue, or the spheroidal and disk-like galaxies
based on their location in the luminosity--redshift space.

\item The 1D and 2D methods show the evident bimodality both of the
$UBV$ rest-frame colour and S\'ersic index distribution up to redshift
\mbox{$z=1$}.

\item The combination of the $UBV$ rest-frame colour and S\'ersic index~$n$
as a function of redshift and luminosity leads to a precise statistical
description of the structure of galaxies and their evolution.
Our method of analysis connects four galaxy parameters, i.e. $UBV$
colour, S\'ersic index, luminosity and redshift, and allows to present
the coevolution of the properties of galaxies belonging to the early-
and late-type classes together with their evolution. 

\item We find that both early- and late-type galaxies simultaneously
become redder and more concentrated both with cosmic time and increasing
luminosity (stellar mass). Early type galaxies, however, display only a
slow change of their concentrations since $z\sim1$; it is already
established by $z\sim1$ and depends much more strongly on their
luminosities. In contrast, late-type galaxies get clearly more
concentrated with cosmic time since $z\sim1$, with only little evolution
in colour, which remains dependent mainly on their luminosity. This
flipped luminosity (mass) and redshift dependence likely reflects
different evolutionary tracks of early- and late-type galaxies before
and after $z\sim1$.

\end{itemize}

We demonstrated that the method presented in this paper is an
improvement manner to separate early- and late-type galaxies, and to
study how their colour and morphology depend on luminosity and
redshift.
This can be used in further investigation of galaxy evolution.


\begin{acknowledgements}
We acknowledge the crucial contribution of the ESO staff for the
management of service observations. In particular, we are deeply
grateful to M. Hilker for his constant help and support of this
program. Italian participation to VIPERS has been funded by INAF
through PRIN 2008, 2010 and 2014 programs. LG and BRG acknowledge
support of the European Research Council through the Darklight ERC
Advanced Research Grant (\# 291521). OLF acknowledges support of the
European Research Council through the EARLY ERC Advanced Research
Grant (\# 268107). AP, KM, and JK have been supported by the National
Science Centre (grants UMO-2012/07/B/ST9/04425 and
UMO-2013/09/D/ST9/04030), the Polish-Swiss Astro Project (co-financed
by a grant from Switzerland, through the Swiss Contribution to the
enlarged European Union). RT acknowledge financial support from the
European Research Council under the European Community's Seventh
Framework Programme (FP7/2007-2013)/ERC grant agreement n. 202686. EB,
FM and LM acknowledge the support from grants ASI-INAF I/023/12/0 and
PRIN MIUR 2010-2011. LM also acknowledges financial support from PRIN
INAF 2012. Research conducted within the scope of the HECOLS
International Associated Laboratory, supported in part by the Polish
NCN grant DEC-2013/08/M/ST9/00664.
\end{acknowledgements}


\bibliographystyle{aa} 
\bibliography{vipers_krywult}

\appendix

\section{Tests of the GALFIT results}
\label{appendix_tests}

\begin{figure}
\centering
\includegraphics[width=\hsize]{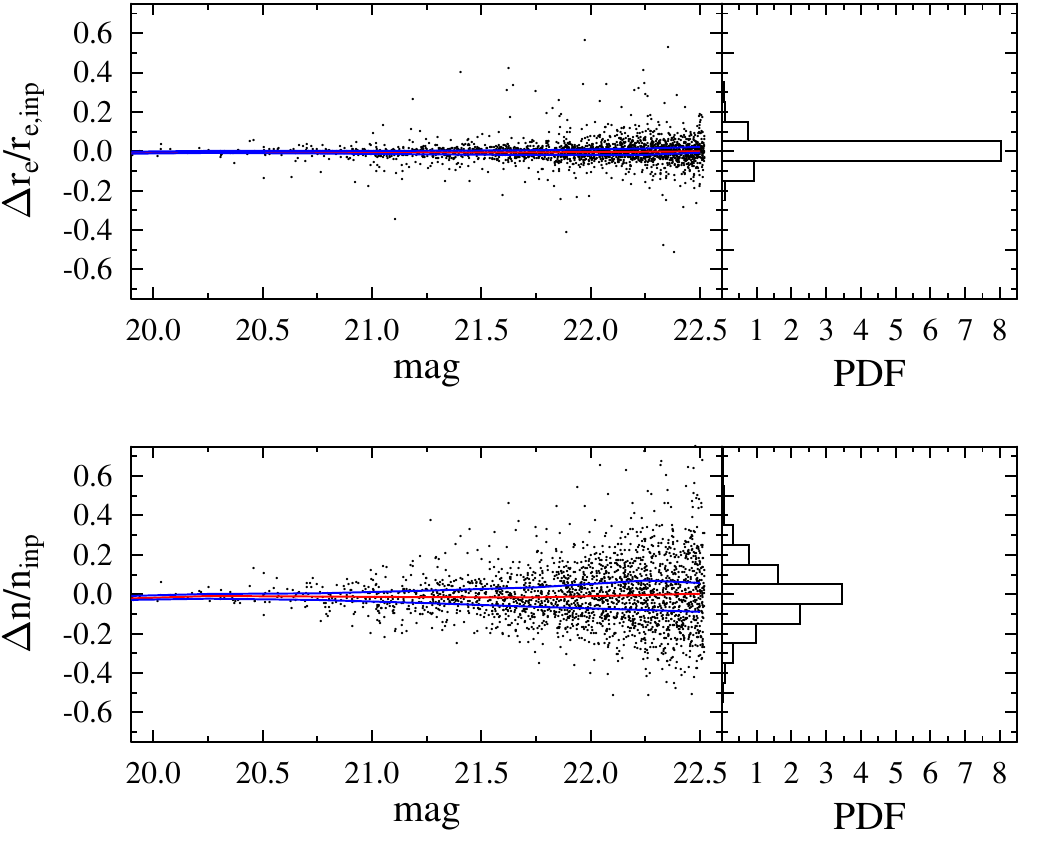}
\caption{Distribution of the fractional deviation of the S\'ersic index $n$
and the effective radius $r_e$ as a function of the apparent magnitude.
{\tt test\_n\_re\_2}}
\label{test_n_re_2}
\end{figure}
\begin{figure*}
\centering
\includegraphics[width=\hsize]{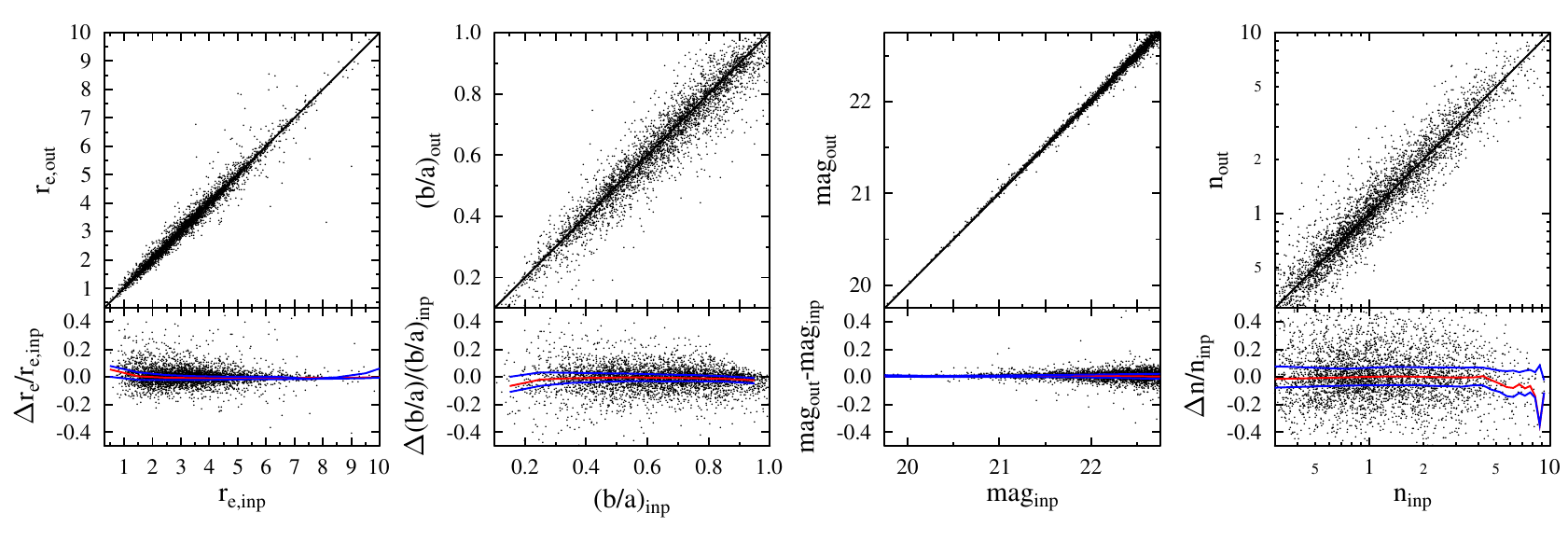}
\caption{The comparison between S\'ersic parameters of $\sim4000$ 
simulated galaxies and their recovered values. Bottom plots show the
fractional deviation of the parameters as a function of the half-light
radius $r_e$ in pixels, minor to major axis ratio $b/a$, apparent magnitude $m$ and S\'ersic index $n$. The red line shows the median whereas the blue
line denotes the $1\sigma$ scatter around the median, defined to enclose 68\% of the points at a given input value.}
\label{galfit_tests1}
\end{figure*}

To assess the robustness of the presented galaxy profile fitting
procedure we perform simulations similar to ones presented in the
literature
\citep{Haussler2007,Longhetti2007,Guo2009,Pannella2009,Mosleh2013}.
To estimate the accuracy of the results obtained from GALFIT, we
applied exactly the same fitting procedure as that used for the real
objects to a set of $\sim4000$ (i.e. $\sim$10\% of a real sample)
artificial galaxies.

The simulated objects were generated using the S\'ersic parameters
from the GALFIT output of the randomly selected VIPERS galaxies.  This
way gives more realistic shape of the galaxy light profile than the
randomly generated parameters of the S\'ersic parameters and allows us
to compare both results for each single object.

Each simulated profile was added to a different background image.
To construct the background we applied the method similar to that
proposed by~\citet{Longhetti2007}.
The background image has been obtained by mosaicing different portions
of the object-free regions of the CFHTLS tile into one large image.
Then the generated profile of each galaxy was superimposed on the
randomly selected region of the background.
The advantage of this method is that the background retains the same
noise characteristic as the real CCD image.
Finally the galaxy profile was convolved with the PSF of the shape
generated at the galaxy position.

Simulated images of galaxies prepared in this way were analysed by
SExtractor and GALFIT using the same procedure as done for the real
objects. The results of our tests are presented in
Fig.~\ref{galfit_tests1}. The top row shows the relation between the
input and output parameter values, whereas the bottom one presents the
estimated uncertainty of the S\'ersic profile parameters, i.e. the
half-light radius $r_e$ in pixels, S\'ersic index $n$, axis ratio $b/a$
and apparent magnitude $m$. The red lines in the bottom figures show
the median and the blue lines indicate the $1\sigma$ scatter around the
median, defined as that which encloses 68\% of the points.

The tests results presented in Fig.~\ref{galfit_tests1} show that the
galaxy apparent magnitude $m$ is the best recovered S\'ersic function
parameter in the simulation and the accuracy of its value decreases
for the less luminous galaxies.  Similar results from GIM2D and GALFIT
were obtained by~\cite{Pannella2009}.

Simulations show that the error of recovered value of the half-light
radius $r_e$ is larger for the smaller galaxies.  However, even in
this case the difference between input and output parameters presented
in Fig~\ref{galfit_tests1} is less than $10\%$.  The small systematic
differences between those values was also reported in other studies
~\citep{Haussler2007, Longhetti2007, Guo2009}.  Moreover, we found
that the axis ratio $b/a$ of the galaxy light profile is robust.  The
uncertainty of this parameter is in the order of a few percent.

The analysis shows that the S\'ersic index $n$ is also well recovered.
However, in this case we observe a larger fractional deviation scatter
around the median than for mentioned previous two parameters.
Figure~\ref{galfit_tests1} shows that the fractional deviation of the
S\'ersic index $n$ is almost uniformly distributed over the $n$.
Similar scatter of the reconstructed parameters is also
reported in other studies~\citep{Haussler2007, Longhetti2007, Guo2009}.

The carried out simulations show a good agreement between the input
and output S\'ersic function parameters.  The $1\sigma$ deviation of
values (i.e. containing 68\% of points) about the parameter median is
narrow, and for majority of the tested objects the error of recovered
parameters are less than 10\%.

\begin{figure*}
\centering
\includegraphics[width=\hsize]{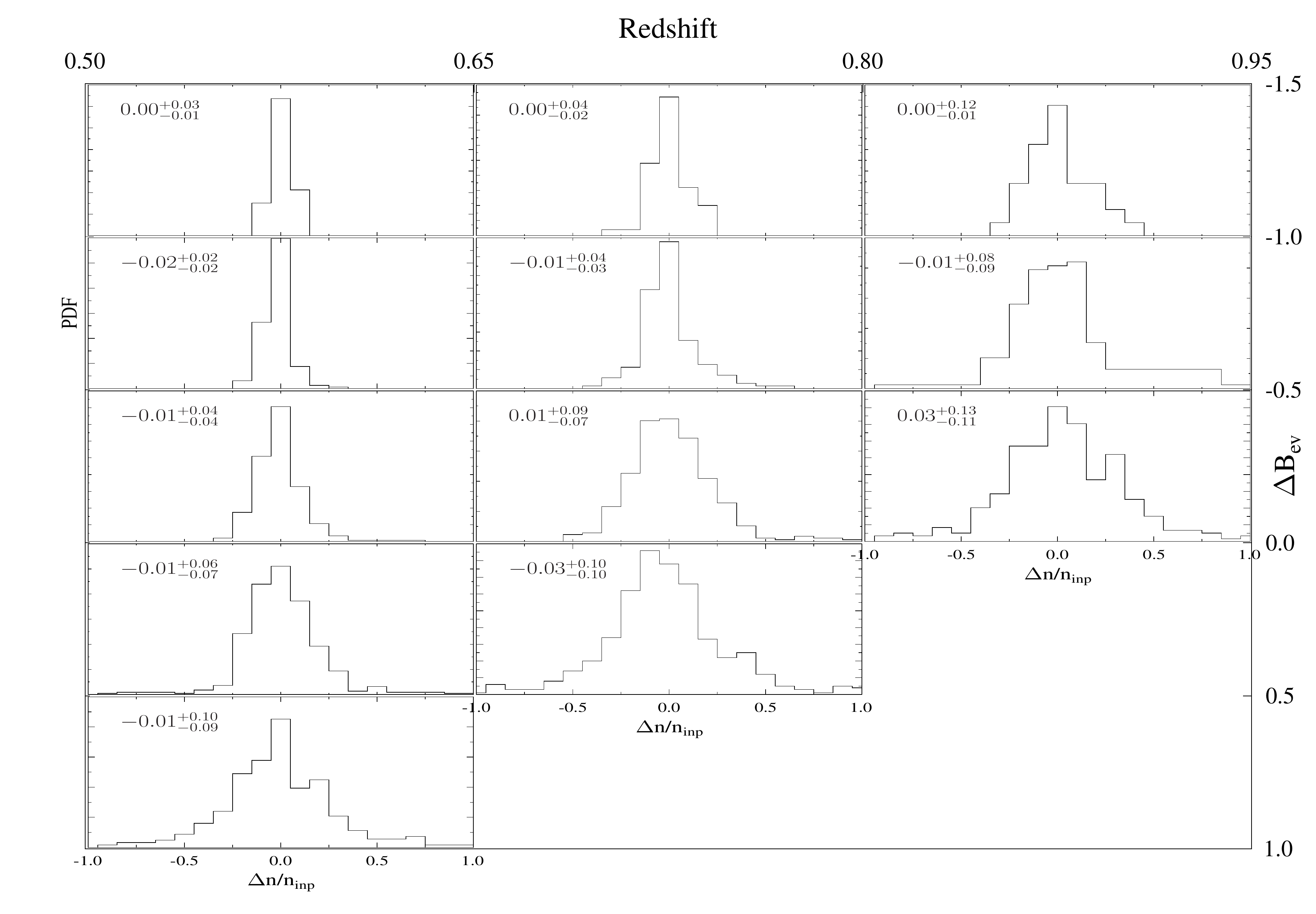}
\caption{Distribution of the fractional deviation of the S\'ersic index
$n$ as a function of redshift and luminosity.
In each bin the histogram of the fractional difference $\Delta n/n_{\rm
inp}$ with their median and $1\sigma$ uncertainties around the median
(i.e. the 16th and 84th percentile of values) are shown.}
\label{n_test_grid}
\end{figure*}

In the study of the distant galaxies the typical size of galaxy
registered on the CCD images is small and can influence their
estimated light profile parameters.  In the next tests we verify the
accuracy of the S\'ersic index $n$ and half-light radius $r_e$ as a
function of the apparent magnitude.  The results are shown in
Figure~\ref{test_n_re_2}.  As expected, this figure shows that faint
galaxies exhibit larger random uncertainties in their S\'ersic index
$\Delta n/n_{\rm inp}$ and half-light radius $\Delta r_e/r_{e,\rm inp}$.

Figure~\ref{test_n_re_2} shows that in both cases the error of
recovered parameters is smaller for the luminous galaxies and as
expected systematically increases for faint objects. For faint
galaxies the external part of the objects can fall under the sky
surface brightness and this effect can lead to increases of the value of
S\'ersic index $n$. The test shows no systematic. The
distribution of errors in the whole analysed luminosity range
is symmetric.

The last test we present shows the fractional difference of the
S\'ersic index as a function of redshift and luminosity.  To do this
we applied the same binning as was used in our analysis and presented
in Fig.~\ref{gv_B_grid}, \ref{grid_n} and \ref{ubv_n_dens_grid}, to
estimate the reliability of our study.

The simulated \mbox{$\sim4000$} objects were generated using the
S\'ersic parameters from the GALFIT output as described in the first
test.  Galaxies are then divided into redshift-luminosity samples to
compute mean and standard deviation of the distributions of the
fractional difference $\Delta n/n_{\rm inp}$.

Figure~\ref{n_test_grid} presents the results and shows histograms and
median value with the $\pm 34\%$ scatter around the median.  The
histograms show that the accuracy of the S\'ersic index $n$ estimation
decreases both with redshift and luminosity.
The most accurate value of $n$ we get is for the nearby and most
luminous galaxies.
As expected, faint galaxies exhibit larger random uncertainties in their
S\'ersic index $n$ parameter, what is consistent with the previous test.
The histograms and numerical values presented in Fig.~\ref{n_test_grid}
show no systematic deviation of the S\'ersic index fitted to the galaxy
images.
The tests we presented show that S\'ersic function parameters computed
by GALFIT from CFHTLS CCD images of the VIPERS galaxies are robust.


\section{Modeling of the PSF}
\label{appendix_psf}

The CFHTLS images were obtained with MegaCam at the prime-focus with
wide-field corrector~\citep{Boulade2000}. However, while the corrector
is optimised to produce a uniformly high quality image of the whole
field of view, it also introduces large-scale non-linear geometrical
distortions~\citep{Cuillandre1996}. This effect together with the
seeing significantly disturb the isotropy of the PSF and have to be
corrected before any further measurements are done from the images.

The elliptical Point Spread Function for the CFHTLS images can be
approximated by the \citet{Moffat1969} function
\begin{equation}
I(r)= I_0\left(1+\left(\frac{r}{\alpha}\right)^2\right)^{-\beta},
\end{equation}
where $I_0$ is the central luminosity, $\beta$ is the profile shape
parameter and $\alpha$ is the half-light radius of the profile.

\begin{figure}
\centering
\includegraphics[width=\hsize]{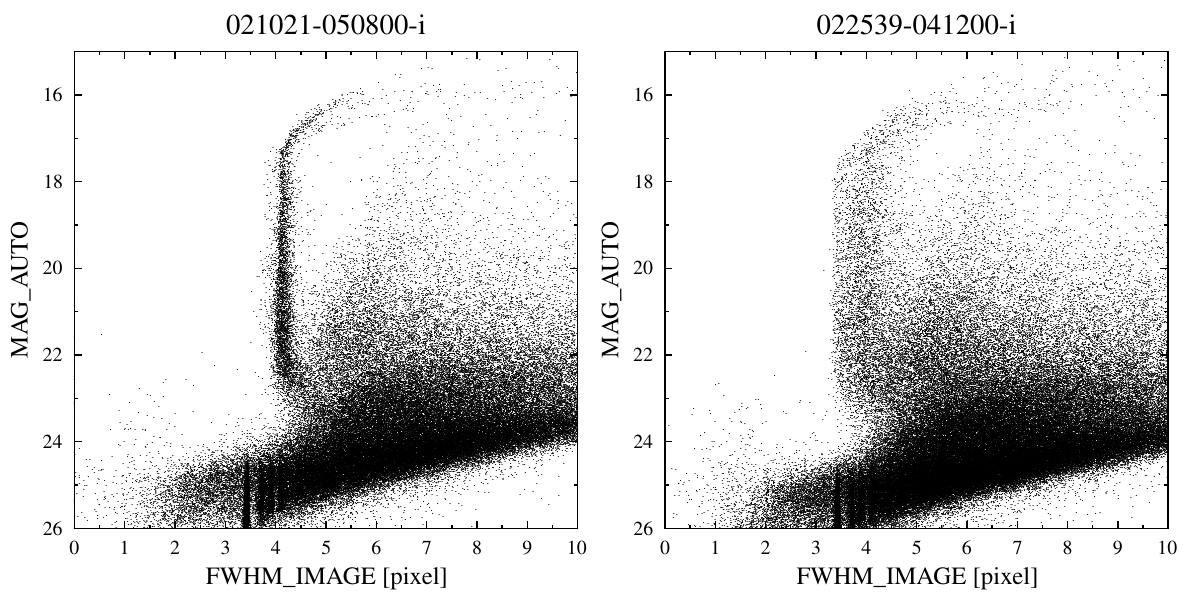}
\caption{MAG\_AUTO versus FWHM\_IMAGE plot used to select PSF
  candidate objects for two CFHTLS tiles with good (left panel) and
  worse (right panel) image quality.}
\label{mag_fwhm}
\end{figure}

To construct a proper PSF for the VIPERS galaxies across the whole
$1\degr\times 1\degr$ CCD field we carefully selected stars from each
CFHTLS tile.
The stars were taken from the stellar branch of the SExtractor
\citep{Bertin1996} MAG\_AUTO versus FWHM\_IMAGE diagram within the
apparent magnitude range from 18 to 22 mag.
Figure~\ref{mag_fwhm} presents two examples of this diagram computed
from images with good and bad quality.
In the first plot the vertical region dominated by the point-like
objects is sharp wheres in the second one it is significanty wider.
To remove small and distorted stars, the objects with SExtractor
ISOAREA\_IMAGE $\le 10$ pixels and ELLIPTICITY $>0.2$ were rejected from
the analysis.
The visual inspection of the CCD images confirmed that these criteria
very well select isolated and non-distorted point-like objects.

\begin{figure}
\centering
\includegraphics[width=\hsize]{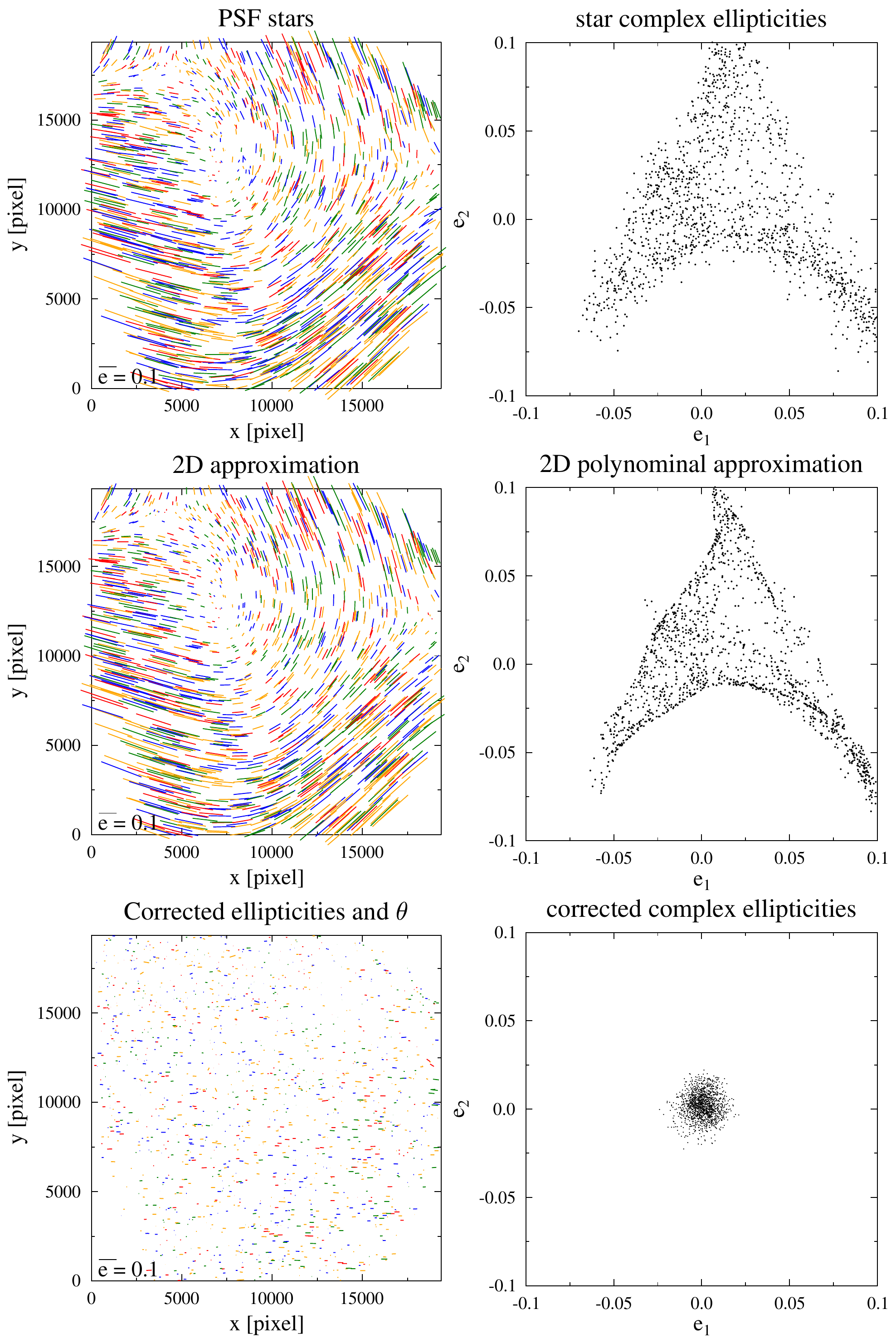}
\caption{Whisker plot (left column) and the complex ellipticities
  (right column) obtained from the PSF parameters for the
  022539-041200 tile in the $i$-band. The upper panels show the
  results obtained from uncorrected star ellipticities, the middle
  panels -- from the 2D polynomial approximation of the Moffat
  parameters and the bottom panels -- the final results after the
  correction for the anisotropy was applied. The small bar at the
  left-bottom corner of all left panels shows the ellipticity $e=0.1$.
  Apparent $i$ magnitudes are marked by colours: values of 18-19 mag
  by red, values of 19-20 mag by green, values of 20-21 mag by blue,
  and values of 21-22 mag by orange.}
\label{whisker_plot}
\end{figure}

The average number of stars used for the approximation of the Moffat
parameters, and uniformly distributed in each of $1\degr\times 1\degr$
CFHTLS tile, is $\sim2000$ and varies from field to field (between
$\sim1000$ and $\sim3500$).

However, the applied method might be somewhat restrictive.
Because of the image distortion presented in some CFHTLS tiles there are
regions where no PSF stars were selected by this algorithm, as shown in
the top-left plot in Fig.~\ref{whisker_plot}.
This occurs mainly in the regions close to the tile border covering
about 2\% of the total VIPERS area.
Since the quality of the PSF plays such an important role in the GALFIT
image deconvolution we excluded galaxies from these regions from the
presented analysis.

In the next step, a Moffat function was fitted to the images of stars
extracted from the CFHTLS tiles in the form of the postage stamps of
size 35x35 pixels each, which is more than 10 times larger than the
FWHM.
Then, for each CFHTLS tile of size of $1\degr\times 1\degr$, the fitted
parameters of the Moffat function were approximated by the
two-dimensional Chebyshev polynomial.
The Chebyshev approximation was used due to its numerical stability and
the smallest maximum deviation from the approximated
function~\citep{Dahlquist1974}.
We have checked polynomials of degree from 5 to 11 and found that
polynominal degree of 7 best approximates the Moffat function parmeters
across the CFHTLS tile.
In this way, we obtained an analytical form of how each of the Moffat
function parameters vary across the whole field, which allowed us to
compute the PSF at the position of every galaxy in the tile.

The first verification of our PSF modeling was performed using the
whisker plot~\citep{VanWaerbeke2000,Tewes2012}. This diagram
demonstrates how the ellipticity $e$ and the orientation $\theta$ of
PSF stars vary across the field. Each selected star is represented by
a line whose length represents the star ellipticity $e$, and is
orientated to match the position angle of the star's major axis. The
colour of each line represent the $i$-band magnitude of the star.  The
first plot in the top row of Figure~\ref{whisker_plot} shows a strong
anisotropy of observed stars which should be corrected for. To do
this, we applied the complex ellipticities \citep{Tewes2012} defined
by
\begin{equation}
e=\frac{\epsilon-1}{\epsilon+1}\exp(i2\theta)=|e|\left(\cos(2\theta) + 
i \sin(2\theta)\right)=e_1+i\cdot e_2,
\end{equation}
where the elongation is defined as $\epsilon = {b/a}$. This
representation is commonly used in anisotropy determination in
gravitational lensing analysis~\citep{Holhjem2009, VanWaerbeke2000}.

We performed a $3\sigma$ clipping procedure on the corrected stellar
complex ellipticities $e_1$ and $e_2$ \citep{Tewes2012}, which removed
most of the stars whose shape was deformed.  After this procedure, for
each tile the parameters of the Moffat function were approximated
again by the two-dimensional Chebyshev polynomial degree of 7.  This
iteration leads us to the final analytical approximation of the PSF
coefficients used to reconstruct the PSF at the position of each
VIPERS galaxy.

As an example of the method described above, Fig.~\ref{whisker_plot}
presents the selected plots obtained from the final Moffat function
parameters for the lower quality CFHTLS\_022539-041200\ tile in
$i-$band.  The diagrams in the second row of the figure were obtained
from our global PSF approximation.  The plots presented in the first
and second rows show very good correlations between the observed and
approximated results.  The last row in Fig.~\ref{whisker_plot} shows
the PSFs of stars after the correction for anisotropy. One can observe
that the field corrected for telescope anisotropy is almost uniform,
which confirms the high quality of our PSF approximation.  Even for a
tile with an image distortion as high as the one presented in this
example, our method accurately maps the variation of the PSF across
the whole field of view. This precise modeling plays a vital role in
the successful galaxy image decomposition by GALFIT.


\end{document}